\documentclass[12pt,preprint]{aastex}

\def\gapprox{\;\rlap{\lower 2.5pt                       
             \hbox{$\sim$}}\raise 1.5pt\hbox{$>$}\;}
\def\lapprox{\;\rlap{\lower 2.5pt                       
             \hbox{$\sim$}}\raise 1.5pt\hbox{$<$}\;}

\shorttitle{{\it FUSE} Magellanic Cloud Extinction Curves}
\shortauthors{Cartledge et al.}

\begin{document}

\title{{\it FUSE} Measurements of Far Ultraviolet Extinction. II. Magellanic
Cloud Sight Lines\footnote{Based on observations with the NASA-CNES-CSA
{\it Far-Ultraviolet Spectroscopic Explorer}, which is operated for NASA by
the Johns Hopkins University under NASA contract NAS-32985.}}

\author{Stefan I. B. Cartledge\altaffilmark{1}, Geoffrey C. Clayton\altaffilmark{1}, Karl D. Gordon\altaffilmark{2}, Brian L. Rachford\altaffilmark{3}, B. T. Draine\altaffilmark{4}, P. G. Martin\altaffilmark{5}, John S. Mathis\altaffilmark{6}, K. A. Misselt\altaffilmark{2}, Ulysses J. Sofia\altaffilmark{7}, D. C. B. Whittet\altaffilmark{8}, and Michael J. Wolff\altaffilmark{9}}

\altaffiltext{1}{Department of Physics and Astronomy, Louisiana State
University, Baton Rouge, LA 70803; scartled@lsu.edu,
gclayton@fenway.phys.lsu.edu}

\altaffiltext{2}{Steward Observatory, University of Arizona, Tucson, AZ 85721;
kgordon@as.arizona.edu, misselt@as.arizona.edu}

\altaffiltext{3}{Center for Astrophysics and Space Astronomy, Department of
Astrophysical and Planetary Sciences, University of Colorado at Boulder, Campus
Box 389, Boulder, CO 80309-0389; rachford@casa.colorado.edu}

\altaffiltext{4}{Princeton University Observatory, Peyton Hall, Princeton, NJ
08544; draine@astro.princeton.edu}

\altaffiltext{5}{Canadian Institute for Theoretical Astrophysics, University of
Toronto, Toronto, Ontario M5S 3H8 Canada; pgmartin@cita.utoronto.ca}

\altaffiltext{6}{University of Wisconsin-Madison, 475 North Charter Street,
Madison, WI 53706; mathis@astro.wisc.edu}

\altaffiltext{7}{Department of Astronomy, Whitman College, Walla Walla, WA
99362; sofiauj@whitman.edu}

\altaffiltext{8}{Department of Physics and Astronomy, Rensselaer Polytechnic
Institute, Troy, NY 12180-3590; whittd@rpi.edu}

\altaffiltext{9}{Space Science Institute, 3100 Marine Street, Suite A353,
Boulder, CO 80303-1058; wolff@colorado.edu}

\begin{abstract}
We present an extinction analysis of nine reddened/comparison star pairs in the
Large and Small Magellanic Clouds based on {\it Far-Ultraviolet Spectroscopic
Explorer (FUSE)} FUV observations. To date, just two LMC sight lines have
probed dust grain composition and size distributions in the Magellanic Clouds
using spectral data for wavelengths as short as 950 {\AA}. We supplement these
two with data from 4 regions distinguished by their IR through UV extinction
curves and grouped as LMCAvg, LMC2, SMC bar and SMC wing. Despite the distinct
characters of extinction in the Clouds and Milky Way, our results are generally
analogous to those found for Galactic curves---namely, that the {\it FUSE}
portions of each extinction curve are described reasonably well by Fitzpatrick
\& Massa curves fitted only to longer wavelength data and lack any dramatic new
extinction features, and any deviations from the Cardelli, Clayton, \& Mathis
(CCM) formalism continue into FUV wavelengths. A Maximum Entropy Method
analysis of all of these curves suggests that LMCAvg and SMC wing sight lines,
whose extinction parameters more closely resemble those for Galactic paths,
require more silicon and/or carbon in dust than current abundance measurements
would indicate are available. The requirements for LMC2 and SMC bar sight lines
do not fully tax the available reservoirs, in part because large grains
contribute less to the extinction in these directions. An intermediate product
of this extinction analysis is the measurement of new H$_2$ abundances in the
Magellanic Clouds. Collectively considering Cloud sight lines that possess
significant H$_2$ column densities, $E(\bv)$/$N$(\ion{H}{1}) ratios are reduced
by significant factors relative to the Galactic mean, whereas the corresponding
$E(\bv)$/$N$(H$_2$) values more closely resemble their Galactic counterpart.
These trends reflect the fact that among these sight lines $f$(H$_2$)-values
are lower than those common in the Milky Way for paths with similar degrees of
reddening.
\end{abstract}


\section{Introduction}
\label{section_introduction}
An important step toward a complete understanding of the formation, structure,
and composition of interstellar dust is the study and discrimination of factors
that produce changes in the observed wavelength-dependent extinction of stars
due to grains. The Large and Small Magellanic Clouds (LMC and SMC, respectively)
are metal-poor relative to the Galaxy; recent estimates imply that the LMC and
SMC have overall metallicities at levels 0.5 and 0.2--0.25, respectively, of
the Galactic ISM \citep{wel01}. Dust-to-gas ratios determined for these
galaxies, as represented by $E(\bv)/N$(\ion{H}{1}), are also reduced (by
factors of about 4 and 10; \citealt{bou85,fit86}), indicating that LMC and SMC
dust components and their formation mechanisms are not inconsistent with
Galactic components and mechanisms. Nevertheless, there are differences in
extinction among these three galaxies.

Early investigations of the extinction properties of Magellanic Cloud dust
identified several sight lines that exhibited marked differences from curves
produced by grains in the Milky Way (e.g., \citealt{nan80,koo81,roc81}). It
was noted in comparing these curves that the 2175 {\AA} bump strength was
reduced in the LMC relative to the Galaxy, and somewhat further diminished
for curves characteristic of the SMC. Similarly, the FUV rise
evident in LMC and SMC extinction curves is stronger than that common in the
Galaxy. Subsequent studies have identified regional variation in Magellanic
Cloud curves, associating a distinctive LMC wavelength dependence with the
supershell LMC2 \citep{mis99}, a steeper UV curve with the star-forming bar
region of the SMC, and extinction more closely resembling Galactic curves with
other portions of the Clouds. Recently, \citet{gor03} (hereafter G03) completed
a comprehensive comparison of Galactic, LMC, and SMC extinction curves from
near-infrared to ultraviolet (UV) wavelengths ($\lambda \gapprox$ 1150
{\AA} or $1/\lambda \lapprox 8.7 \micron^{-1}$); G03 confirmed that curves
characteristic of the LMC2 grouping and the SMC bar do not conform to the
\citet{car89} (CCM) parameterization of Galactic extinction based on $R_V\,
[\equiv A(V)/E(\bv)]$.

The strength of their UV extinction is a key feature distinguishing LMC, SMC,
and Galactic curves; in particular, the steeper short-wavelength slopes
associated with Magellanic Cloud extinction curves can be diagnostic of
differences in dust compositions and grain size distributions, given
appropriate measures of elemental abundances in their respective interstellar
media \citep{dra03}. However, the extension of Magellanic Cloud extinction
curves into the FUV ($>$ 8.7 $\micron^{-1}$) is complicated by the requirement
that interstellar H$_2$ absorption be removed from the spectra before an
extinction curve can be
produced using the pair method. Previous efforts in this area, even for Milky
Way sight lines, have been hampered by a lack of high quality data. In
particular, instrumental issues of scattered light, time-variable sensitivity,
and a limited sample (both target and comparison objects) restricted the
utility of the {\it Copernicus} dataset (cf. \citealt{jen86}; \citealt{sno90}),
the {\it Voyager} UVS data suffer from low resolution which prevents explicit
identification and removal of the molecular hydrogen contribution, and the
remaining available data [e.g., rocket - \citealt{gre92,lew05}; {\it Hopkins
Ultraviolet Telescope (HUT)} - \citealt{bus94}; {\it Orbiting Retrievable Far
and Extreme Ultraviolet Spectrometer (ORFEUS)} - \citealt{sas02}] include only
a few Galactic
sight lines. {\it Far Ultraviolet Spectroscopic Explorer (FUSE)}, however, has
revolutionized our ability to study interstellar molecular hydrogen. Its
combination of slightly better spectral resolution and much greater detector
sensitivity than {\it Copernicus} allows H$_2$ to be detected throughout much
larger and more diverse portions of the Galaxy and even in the Magellanic
Clouds \citep{shu00,rac02,tum02}. Thus, {\it FUSE} data are uniquely suited to
investigating the FUV extinction properties of Milky Way, LMC, and SMC sight
lines and what any differences reveal about the dust populations in each
galaxy.

Recently, \citet{cla03} used a modified maximum entropy method (MEM) to fit
extinction components to IR through UV curves for stars both in the Galaxy and
in the Clouds; they demonstrated that the average SMC bar extinction curve was
best fit using a grain distribution in which small silicate and amorphous
carbon grains play a much larger role than in the mean Galactic extinction.
However, this result is at odds with \citet{wel01}, whose data suggested
negligible silicon depletion in several gas cloud components along a sight line
probing the SMC ISM. In order to address issues such as dust composition,
particularly for small grains enhancing FUV extinction in the Clouds, it would
be helpful to explore LMC and SMC extinction curves shortward of the UV range
that \citet{cla03} were confined to by their {\it International Ultraviolet
Explorer (IUE)} data; unfortunately, FUV extinction curves have been published
to date for only very few reddened Magellanic Cloud stars \citep{cla96,hut01}.

In this paper, we construct new extinction curves for nine pairs of reddened
and unreddened LMC and SMC stars by supplementing the IR through UV curves
published by G03 with recent {\it FUSE} observations of 16 Magellanic Cloud
targets. The {\it FUSE} data allow us to extend the \citet{cla03} analysis of
dust composition and grain size distribution to FUV wavelengths, where
extinction is dominated by small grains and the differences between Galactic
and Magellanic Cloud curves are particularly prominent. This is the second in a
series of three papers exploring extinction in the FUV. Paper I \citep{sof05}
dealt with a small set of Galactic extinction curves characterized by a broad
range of $R_V$-values; Paper III (in preparation) will examine a much larger
sample of Galactic sight lines in an effort to search rigorously for any trends
among observed extinction parameters when FUV data are fully considered.

\section{Observations and Data Extraction} 
\label{section_obsandprocs}
\begin{figure}
\plotone{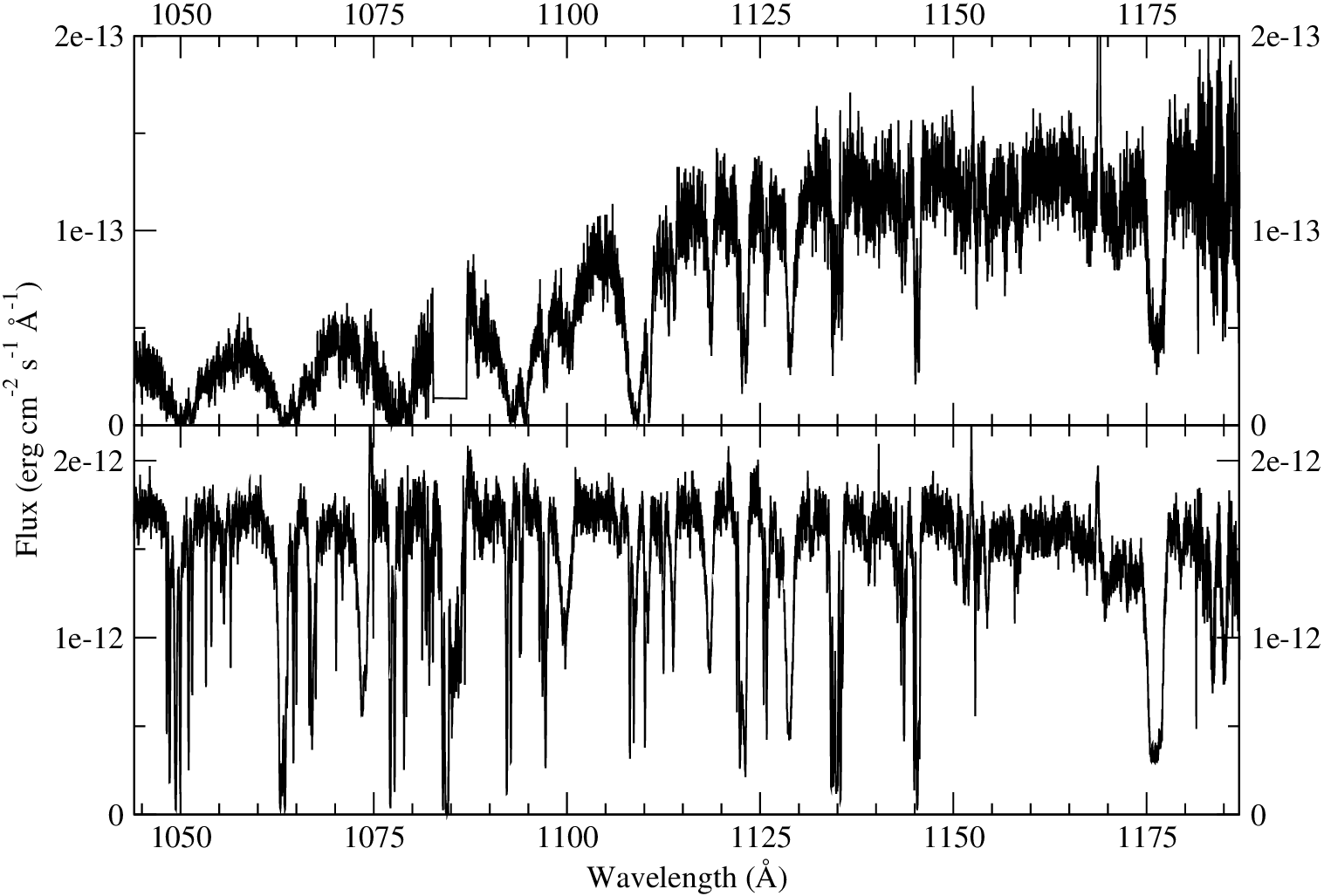}
\caption{Sample {\it FUSE} spectra.\newline
Sample {\it FUSE} spectra are plotted above, including an example of poor
quality data in the upper panel (for AzV 456) and fine quality data in the lower
panel (for AzV 70); these spectra also comprise an extinction pair. Of further
note, the AzV 456 continuum drops dramatically between 1080 and 1120 {\AA},
giving rise to a peculiar kink in the extinction curve discussed later in the
text.\label{spectra}}
\end{figure}

The present sample of 16 LMC and SMC stars includes all stars in the
{\it FUSE} archive for which the spectral match between the reddened and
unreddened stars has been rigorously evaluated and there exist spectral data
from IR to UV wavelengths. Each reddened star has previously been studied by
G03; we use the same photometry sources and procedures to construct the IR to
UV portions of the extinction curves for the pairs adopted in this study,
with the exception that early 2MASS photometry has been superseded in our
analysis by data from the recent {\it All Sky Release}. In particular, it
should be noted that the $R_V$ values listed in Table~\ref{exttab}, which
summarizes the properties of each extinction pair, differ slightly from those
published by G03 due to this update of the IR data. Consequently, the values
of $A_V$ and $N$(\ion{H}{1})/$A_V$ in the table are also affected. The FUV
portions of the extinction curves (shortward of the \ion{H}{1} Ly$\alpha$ line)
were derived from the new and archival {\it FUSE} observations listed in
Table~\ref{obstab}. All of the raw data were processed using the latest release
of CALFUSE (v3.0). The calibrated spectra for each channel were
cross-correlated, shifted to a common wavelength scale, and combined based on
exposure time in cases where several observations of a star existed. Also, in
cases where the different observations or different {\it FUSE} channels
produced conflicting flux levels for a given wavelength (e.g., due to channel
misalignment), the data for these observations or channels were linearly
adjusted to match the maximum observed flux across the regions of overlap.
General
details on {\it FUSE} observations have been published by \citet{sah00a}. We
point out here, however, that each observation produces eight spectra nearly
100 {\AA} in length with individual dispersion characteristics. The LiF channel
spectra (LiF1A, LiF1B, LiF2A, LiF2B) cover most wavelengths from 990--1190
{\AA} twice over; the spectral region from 910--1110 {\AA} are similarly
sampled by the SiC channels (SiC1A, SiC1B, SiC2A, SiC2B). Sample spectra of
both poor and fine quality appear in Figure~\ref{spectra}, using the data from
the reddened star AzV 456 and its comparison star AzV 70, respectively.

\subsection{Molecular Hydrogen Modelling and Removal}
\label{ssection_h2modelling}

Strong FUV H$_2$ absorption bands can significantly alter the appearance of
stellar spectra, depending upon the column density of molecular gas along the
line of sight. Consequently, a molecular hydrogen model was constructed for
each sight line using procedures patterned generally after those used by
\citet{rac01,rac02}. The underlying assumption we made in measuring the
interstellar H$_2$ for each sight line was that the model includes only two
velocity components, one corresponding to Milky Way gas and one associated with
H$_2$ in the appropriate Magellanic Cloud. The equivalent width measurements
for distinct profiles of each component were entered into separate tables, so
that molecular hydrogen measurements could be made individually for each galaxy
using a curve-of-growth (CoG) analysis.
Unfortunately, the velocity separation of Galactic and Magellanic Cloud
components often resulted in blended absorption profiles at several points in a
given spectrum. In general, the construction of a reasonable CoG for most of
the sight lines required more equivalent width measurements than were available
from the unblended and isolated H$_2$ absorption lines. To remedy this problem,
the profile-fitting code FITS6P\footnote{The FITS6P code models absorption
profiles based on varying the column densities, $b$-values, and velocities of
input interstellar components. More details on the algorithm can be found in
\citet{wel91}.} and IRAF\footnote{IRAF is distributed by the National Optical
Astronomy Observatories, which are operated by the Association of Universities
for Research in Astronomy, Inc., under cooperative agreement with the National
Science Foundation.} plotting routines were used in conjunction with our
IDL\footnote{IDL is an acronym for Interactive Data Language, a common
programming tool developed by Research Systems, Inc.} code \citep{rac01,rac02}
to deconvolve blended profiles and fill out each CoG with as many measurements
as could reliably be made. The CoG analysis, however, was generally limited to
$J \geq 3$ because the $J \leq 2$ lines were saturated to the extent that they
possessed broad damping wings that the equivalent width measurement algorithm
could not reliably distinguish from the stellar continuum. Thus, an iterative
curve-fitting procedure
was an additional requirement so that simultaneous determinations of Galactic
and Magellanic Cloud column densities for these lower H$_2$ rotational
excitation levels could be made. The complete molecular hydrogen models,
including column densities for the available $J$ levels, the associated
$b$-value, and the derived kinetic and excitation temperatures (these last
properties are discussed in section~\ref{ssection_h2}), are summarized in
Tables~\ref{MWmolhydtab} and \ref{MCmolhydtab} for all of our LMC and SMC sight
lines; Table~\ref{MWmolhydtab} characterizes the Galactic component for each
sight line where one was detected, while Table~\ref{MCmolhydtab} details the
properties of each Magellanic Cloud component.

Once molecular hydrogen absorption models had been constructed, each spectrum
was corrected for this absorption. All of the {\it FUSE} data for each star
were combined into a single spectrum by shifting the calibrated channel spectra
to a common wavelength reference and averaging the fluxes, weighted by the flux
uncertainty at each point in the overlapping regions. Although there are
significant differences in the data dispersion properties for each channel, the
merged spectrum for each star is of sufficient quality for the construction of
a reliable extinction curve. The ``worm'' problem in {\it FUSE} data
\citep{sah00b} was strong only in LiF1B. Its appearance in the data for this
detector segment was identified by a comparison with the LiF2A flux as the data
for all detector segments were merged into a single spectrum for each sight
line; the portions of the LiF1B spectrum contaminated by the worm were eliminated from the merging process. A running cross-correlation of the normalized
H$_2$ absorption profile corrected for any wavelength mismatches, and the full
final model was divided into the merged {\it FUSE} spectrum to complete the
``removal'' of the molecular hydrogen features. Atomic hydrogen was not
measured for each sight line individually, but its signature, specifically the
Ly$\alpha$ and Ly$\beta$ lines, was removed after the ratio of the reddened and
comparison spectra was calculated as a function of wavelength. Although {\it
FUSE} spectra cover wavelengths from 905 through 1187 {\AA}, data shortward of
1044 {\AA} (our Ly$\beta$ cutoff) were not included in this analysis due to
limitations in data quality.

In the interest of consistency in instrument response, {\it FUSE} flux levels
were then manually rescaled to match {\it IUE} fluxes over the wavelength range
common to both instruments (1150 $-$ 1185 {\AA}) when the full IR to FUV
spectra were constructed; given recent improvements to the calibration of {\it
IUE} data \citep{mas00}, flux errors associated with this instrument have not
yet been surpassed by the latest {\it FUSE} calibration \citep{sah00a}.
Generally, mean fluxes for the data from each instrument agreed to within 20\%,
although there were a few targets for which the discrepancies exceeded this
range. The scalings applied to each {\it FUSE} spectrum (the ratio of {\it IUE}
to {\it FUSE} flux in the overlap region) are given in Table~\ref{fluxshifts}.

\subsection{The Extinction Curves}
\label{ssection_curves} 
We have constructed extinction curves using the pair method (e.g.,
\citealt{mas83}), normalized to $A(V)$ through $R_V$; the pairs of reddened and
unreddened stars in the LMC and SMC were selected (G03) and are listed in
Table~\ref{exttab}. In an effort to minimize uncertainties in each extinction
curve, particularly from spectral mismatch, a detailed comparison of the {\it
FUSE} spectrum for each star with similarly-typed unreddened candidates was
performed, and the extinction pairs matched by G03 using {\it IUE} were also
found to be well-matched in the FUV. Assembly of the extinction curve
corresponding to each pair was accomplished using the
same procedures and near-infrared through UV data outlined by G03. We have also
constructed average extinction curves for the subsets of the current sample
associated with the LMCAvg and LMC2 regions considered previously in the UV
(\citealt{mis99}; G03). Each of the individual LMC curves and
the average curves have been fit to the \citet{fit90} parameterization; the
results are presented in Table~\ref{fmtab}. The two sight lines in the SMC, one
representing each of the bar and wing regions, are considered individually.

\begin{figure}
\plotone{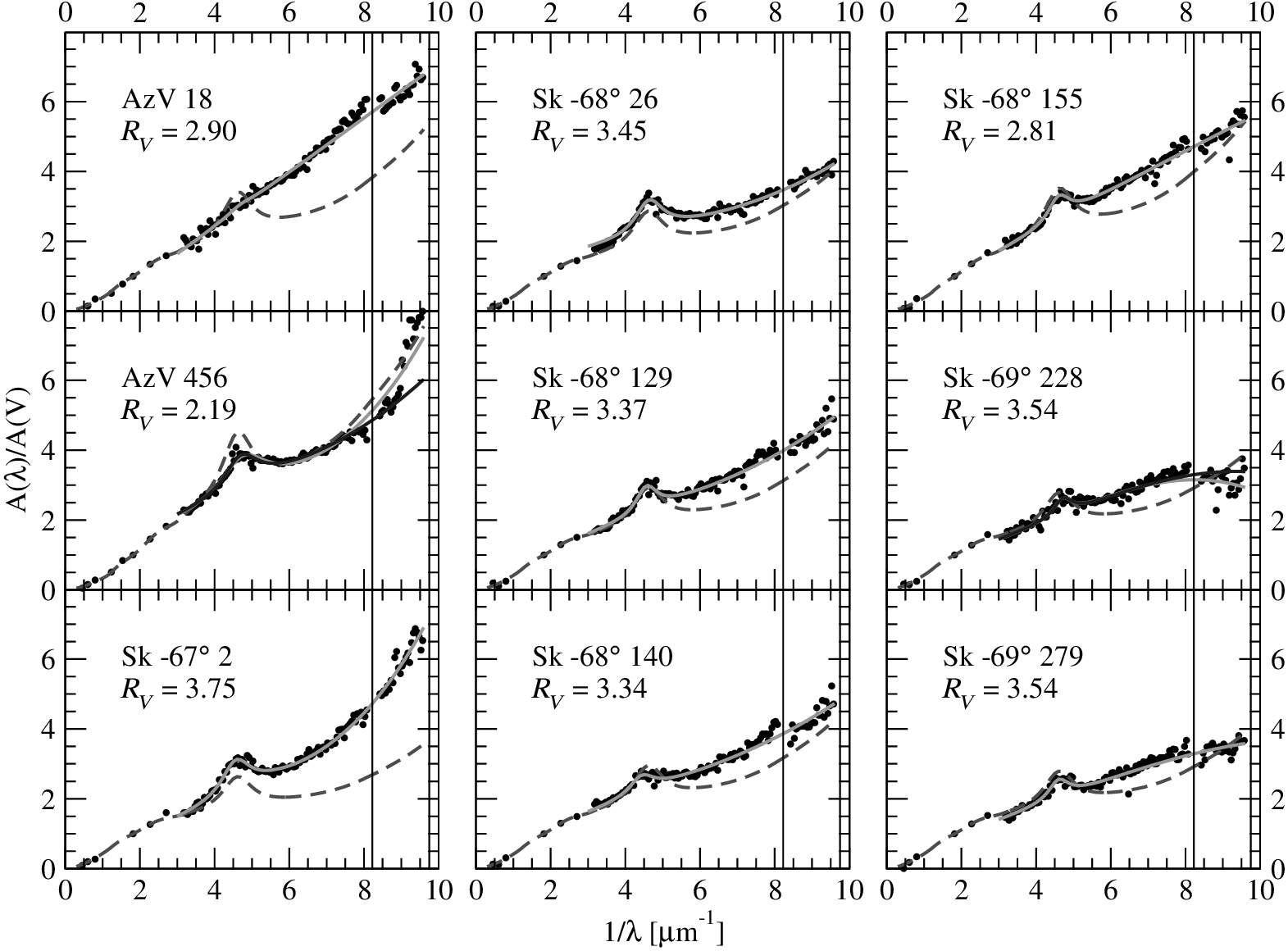}
\caption{Magellanic Cloud Extinction Curves.\newline
The extinction curves for each of the reddened/unreddened star pairs in
Table~\ref{exttab} are plotted above; IR, optical, UV (from {\it IUE}), and FUV
(from {\it FUSE}) data are all included. A CCM curve based on the $R_V$ value
derived from the IR and optical portions of each extinction curve is also
plotted using a dashed line, and the FM fit to the full curves are indicated in
each panel by solid grey lines. In each case but AzV 456 and Sk -69\degr228,
the FM fit for only IR through UV data almost precisely overlaps the FM fit to
the full curve. For these two exceptions, the IR-to-UV fit is represented by a
solid dark line. The vertical lines identify central wavelengths for Ly$\alpha$
and Ly$\beta$.\label{allcurves}}
\end{figure}

\begin{figure}
\epsscale{0.75}
\plotone{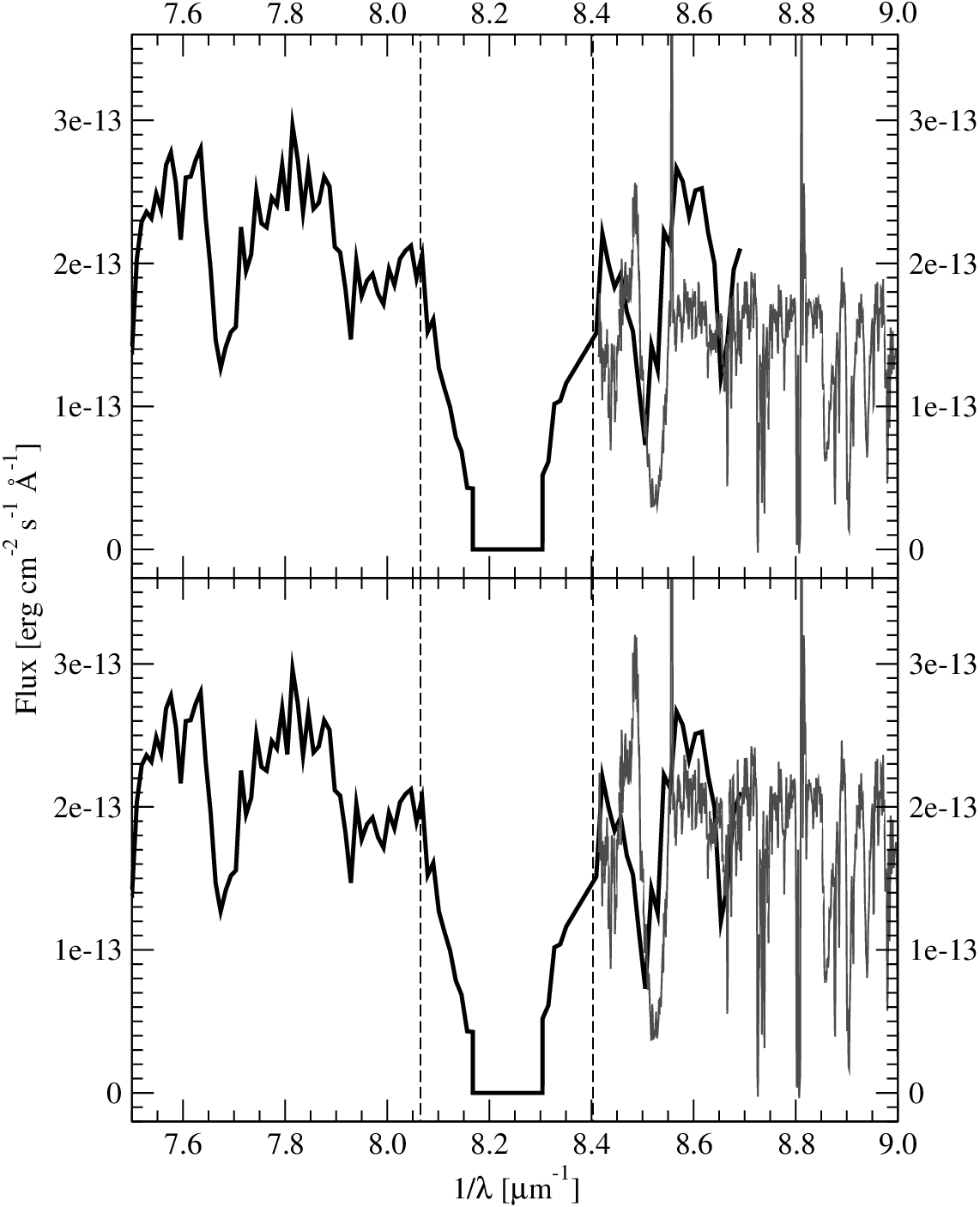}
\caption{Sk -69\degr228 Spectral Match.\newline
{\it IUE} and {\it FUSE} spectra in a region that includes these
instruments' overlapping wavelength coverage have been plotted above for the
star Sk -69\degr228; {\it IUE} fluxes are represented by the heavy solid
line and the light grey line signifies {\it FUSE} data. The top and bottom
panels depict the flux levels before and after the linear corrections were
applied. These corrections generally reduced the size of extinction curve
offsets between UV and FUV data. The vertical dashed lines delimit portions of
the spectrum contaminated by Ly$\alpha$ absorption that have been eliminated
from the curves plotted in Figure~\ref{allcurves}; the central portions have
also been set to zero in this plot.\label{fluxmatch}}
\end{figure}

A quick examination of the individual Magellanic Cloud extinction curves
presented in Figure~\ref{allcurves} reveals that the {\it FUSE} portion of each
curve is generally consistent with an extension of the data from longer
wavelengths. When the calibrated and H$_2$-adjusted fluxes were initially
compared, however, slight offsets were apparent between the portions of the
curve on either side of the Ly$\alpha$ line. Comparing the size of the gaps in
the Magellanic Cloud data with the Galactic curves studied in Paper I led to
the conclusion that the quality of the data for our reddened stars might have
compromised the accuracy of the extinction curve construction. In particular,
the S/N of the {\it FUSE} data seemed to be inversely proportional to the size
of the offset and there were noticeable flux level differences for some spectra
between the {\it IUE} and the merged {\it FUSE} data in the spectral overlap
region. The
Galactic sight lines had much better S/N values, negligible offsets, and good
agreement between {\it IUE} and {\it FUSE} fluxes, whereas the Magellanic Cloud
data suffered generally poor S/N, noticeable offsets, and sometimes poor flux
agreement in the overlap. In order to reduce the offsets, the {\it FUSE} data
for the Cloud sight lines were linearly adjusted (a wavelength-independent
multiplicative factor was applied to the flux) over the entire FUV bandpass
so that their fluxes matched {\it IUE} in the instruments' overlap region. An
example of one of the more egregious cases is plotted in
Figure~\ref{fluxmatch}, including a comparison of the {\it FUSE} flux levels
both before and after adjustment to the {\it IUE} flux level. The magnitude of
each shift was estimated in increments of $0.05 \times flux_{\rm IUE}$; the
systematic uncertainty associated with this adjustment has been propagated
through all subsequent error calculations. Typically, the shifts improved the
appearance of the curves, although the offsets were not eliminated in all
cases. Given the general smoothness of the transition between UV and FUV
portions of the extinction curve in most cases, it might seem reasonable to
introduce further shifts to align these segments; nevertheless, because the
observed offsets fall generally within the error bounds, because we are
interested in the detailed shape of the curve at UV and FUV wavelengths, and
since we have no independent objective basis for such shifts, no further
processing of the individual curves has been performed. It should also be noted
that any residual offsets appear to be biased in the direction of decreased
$A(\lambda)/A(V)$; thus, abundance requirements derived from the FUV data may
be slight underestimates.

\section{Discussion}
\label{section_discussion}

\subsection{H$_2$ in the Magellanic Clouds}
\label{ssection_h2}
As already mentioned, a notable difficulty in deriving extinction curves that
extend into the FUV is the existence of strong molecular hydrogen absorption
bands longward of the Ly$\alpha$ line. Nevertheless, stellar observations of
sufficient quality to allow the H$_2$ absorption to be modelled and removed not
only permit the construction of an extinction curve but also provide more direct
information on the intervening ISM. Recently, \citet{tum02} completed a survey
of molecular hydrogen in the LMC and SMC using several {\it FUSE} observations
of each Cloud; our sample introduces 15 new sight lines into the mix.

In considering these new molecular hydrogen measurements, however, it should be
recognized that the current {\it FUSE} data generally have much lower
signal-to-noise ratios than Galactic observations. In fact, they are also
somewhat lower than the values for previously-observed Magellanic Cloud targets.
\citet{tum02} reported typical 4$\sigma$ equivalent width limits of 30--40
m{\AA}, a level similar to values of 20--50 m{\AA} for our comparison stars but
much lower than the 50--120 m{\AA} characteristic of the reddened targets.
Because of these data limitations, and because the goal of generating a
reliable FUV continuum was given a somewhat higher priority than making a
rigorous assessment of molecular hydrogen, any column densities smaller than
about 10$^{17}$ cm$^{-2}$, and a few that are somewhat larger, are subject to
sizable uncertainties. Nevertheless, we consider the models listed in
Tables~\ref{MWmolhydtab} and \ref{MCmolhydtab} to be reasonable tallies of the
H$_2$ along each sight line.

\begin{figure}
\plotone{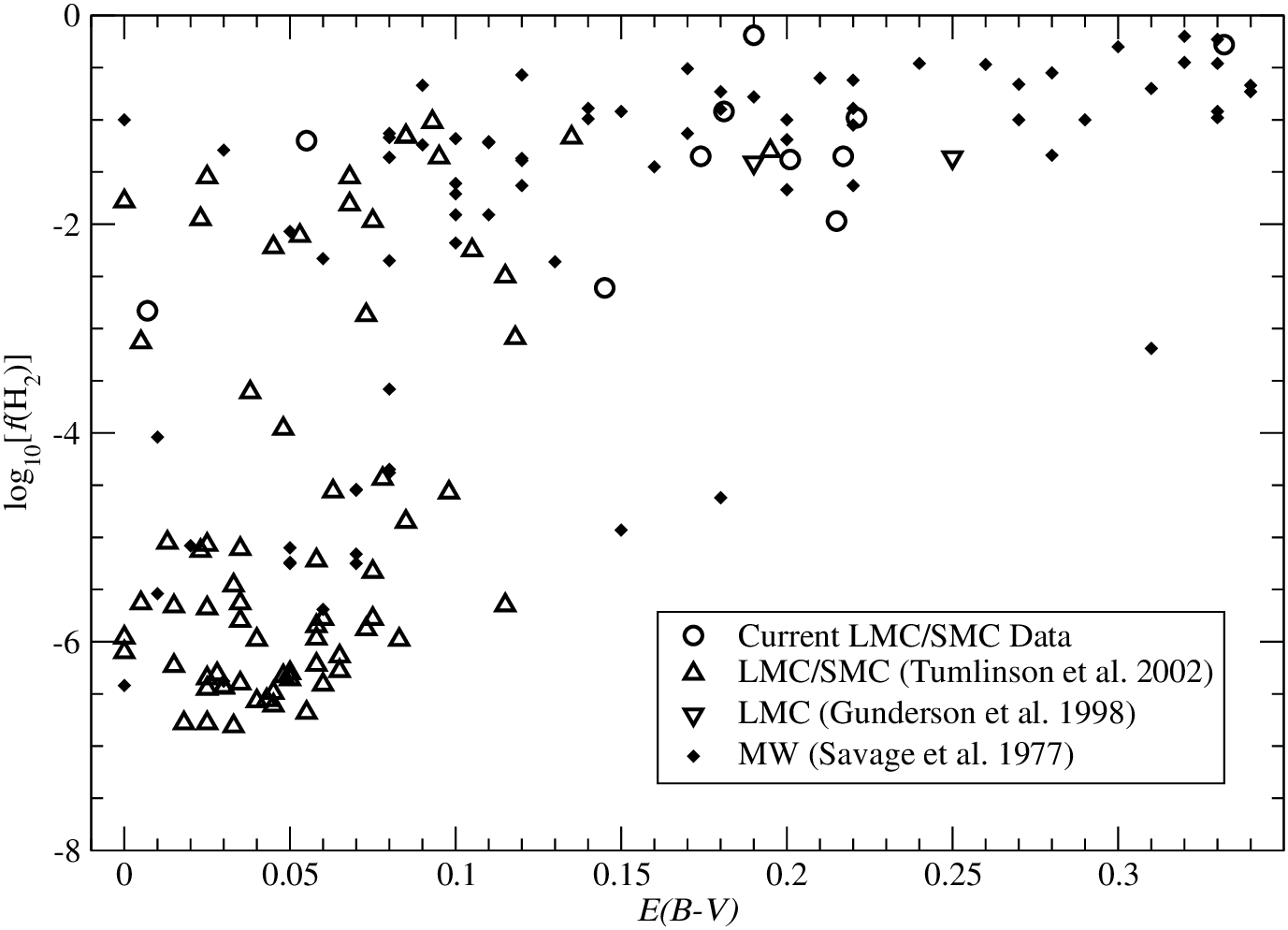}
\caption{Galactic and Magellanic Cloud Molecular Hydrogen Fractions.\newline
The variation of molecular hydrogen fraction $f$(H$_2$) with color excess
$E(\bv)$ is plotted above for sight lines passing through material associated
with the Magellanic Clouds and the Milky Way. The current sample includes
stars that are more heavily reddened that the \citet{tum02} sample and that
emphasize the similarities between how $f$(H$_2$) relates to $E(\bv)$ in these
galaxies. The label $E(\bv)$ used in this caption and for the $x$-axis in the
plot refer to unadjusted values in the Galaxy but $E^\prime(\bv)$ for
Magellanic Cloud sight lines.
\label{f(H2)comp}}
\end{figure}

In order to more fully understand how the new sight lines mesh with the
\citet{tum02} Magellanic Cloud H$_2$ analysis, we have adopted the atomic
hydrogen
measurements previously published by \citet{fit85a,fit85b} for the individual
sight lines (see Table~\ref{hydtab}); for the pair listings in the table, we
use the values derived directly from the extinction curves by G03. It should be
noted here that the process of determining atomic hydrogen column densities for
the Magellanic Clouds is complicated by the large quantities detected along
each sight line. The breadth of the Ly$\alpha$ absorption profile produced by
typical interstellar hydrogen distributions with line-of-sight lengths
exceeding even a few kiloparsecs is sufficient to mask distinctions between
separate velocity components; in particular, the profiles for atomic hydrogen
gas in the LMC or SMC are often inextricably blended with their Galactic
counterparts. \citet{tum02} dealt with this problem by calibrating 21 cm
emission measurements to Ly$\alpha$ from the sight lines in their sample for
which the Galactic and Magellanic Cloud 1215 {\AA} profiles were not severely
blended. Atomic hydrogen column densities for the remaining paths in the data
set were then estimated using this calibration and the observed 21 cm emission.
G03 assessed Magellanic Cloud \ion{H}{1} column densities by measuring the
Ly$\alpha$ profile in the reddened-to-comparison ratio spectrum for each
extinction pair, under the assumption that the Milky Way components would
cancel each other. Each of these methods is subject to a significant level of
uncertainty; if vertical error bars were to be plotted in
Figure~\ref{f(H2)comp}, among the Magellanic Cloud sight lines they might
typically be about 0.3 dex and occasionally larger than 1.0 dex. Nevertheless,
differences between the \citet{fit85a,fit85b} \ion{H}{1} values for the
individual sight lines in each extinction pair generally are well matched by
the corresponding column densities derived by G03.

Considered individually, our LMC and SMC sight lines and those studied by
\citet{tum02} exhibit similar ranges in characteristics such as the molecular
hydrogen fraction $f$(H$_2$) ($\equiv 2N({\rm H}_2)/[N($\ion{H}{1}$) + 2N({\rm
H}_2)]$) and the Magellanic Cloud portion of the color excess
$E^\prime(\bv)$;\footnote{Following the procedure of \citet{tum02}, we adopt
the definition $E^\prime(\bv)$ = $E(\bv) - x$, where $x$ is 0.075 and 0.037 for
LMC and SMC paths, respectively.} values for the two LMC sight lines studied by
\citet{gun98} also match these samples. Since the current dataset was selected
for its extinction properties, we cannot speak to the overall frequency of
molecular hydrogen detection in the Clouds. However, as shown in
Figure~\ref{f(H2)comp}, the $f$(H$_2$)-values for paths along which we detect a
Cloud-based H$_2$ component agree with the levels set by previously observed
LMC and Galactic sight lines of large $E(\bv)$. Notably, the sight lines toward
AzV 456 and Sk $-$67\degr2 possess larger molecular hydrogen fractions than any
examined by \citet{tum02}; meanwhile, among the paths directed toward our
comparison stars we detect no Magellanic H$_2$ components.

Based on the previously-studied sight lines toward Sk -66\degr19 and Sk
-69\degr270, it was found that while the gas-to-dust ratio
$N$(\ion{H}{1})/$E^\prime(\bv)$ is much larger in the Clouds than typically
seen in the Milky Way, the quantity $N$(H$_2$)/$E^\prime(\bv)$ has a similar
value in each galaxy (\citealt{cla96}; Gunderson et al. 1998). This suggests a
close relationship between dust and $H_2$. Similar behavior is evident for our
new sight lines, since although the ratios $N$(\ion{H}{1})/$E^\prime(\bv)$ for
these paths are roughly 4--10 times larger than the Galactic mean, comparing
H$_2$ column densities with the same color excesses leads to a range of values
around the Galactic average. In this way our new sight lines complement the
paths studied by Gunderson et al. (1998) and the more heavily reddened paths of
\citet{tum02}, implying generally lower values for $f$(H$_2$) in the
Clouds; nevertheless, the current dataset includes sight lines with unique
properties. Plotting the LMC $N$(\ion{H}{1})/$E^\prime(\bv)$ ratio as a
function of $E(B-V)$ (see Figure~\ref{gtdebvplot}), there is a clear
distinction between Sk -67\degr2 and other reddened sight lines: this one LMC
path is characterized by an ``atomic'' ratio appropriate to Galactic curves,
but because of its large molecular hydrogen column density the total
gas-to-dust ratio approaches the values for other LMC sight lines. Coupled with
its Milky Way-like $N$(\ion{H}{1}) ratio, the agreement exhibited between the
gas-to-dust ratios for Sk -67\degr2, the sight line with the largest
$f$(H$_2$)-value in our sample, and sight lines dominated by atomic gas
reinforces the concept that the mechanisms governing dust formation in the
Galaxy and LMC are similar and that any differences in extinction arise out of
how interstellar environmental conditions are manifested in dust grain
population characteristics. The SMC sight line AzV 456 also possesses a
``Galactic'' value for $N$(\ion{H}{1})/$E^\prime(\bv)$ that gives a $N{\rm
(H_{Total})/}E^\prime(\bv)$ ratio more closely resembling the values determined
by \citet{tum02} for other SMC paths.

Further details of the interstellar conditions, specifically the temperatures
$T_{01}$ and $T_{ex}$, in the Clouds can be derived from the relative
rotational level populations of the individual sight lines
(Table~\ref{MCmolhydtab}). When the cloud density is sufficiently high and
H$_2$ column densities are large enough for the gas to become substantially
molecular, ortho-para conversion will bring $T_{01}$ ($\equiv \frac{E_1 -
E_0}{k} \frac{\log(e)}{\log(N(J=0)/g_0) - \log(N(1)/g_1)} = \frac{170 {\rm
K}}{\ln(9 N(0)/N(1))}$) close to the kinetic temperature. Except for the AzV
462 sight line, its value is fairly uniform among the LMC and SMC components of
our H$_2$ models and we derive a weighted mean temperature of 50$\pm$3 K. The
Magellanic Cloud component of H$_2$ toward AzV 462 has the lowest column
density among our extragalactic values, and the assumption that this value is
consistent with the kinetic temperature breaks down. The mean temperature we
have derived for the Magellanic Cloud components neglects AzV 462 and is
somewhat lower than the overall Galactic mean of 77$\pm$17 K \citep{sav77}, but
is consistent with both the value 55$\pm$8 K derived from the Galactic
subsample for which $N({\rm H}_2) > 20.4$ \citep{rac02} and temperatures
derived by \citet{tum02} for Magellanic Cloud gas along sight lines with some
of the largest molecular column densities. It should be noted that the errors
ascribed to each temperature listed in Tables~\ref{MWmolhydtab} and
\ref{MCmolhydtab} were derived from column density uncertainties, which were
often large. Thus, kinetic temperatures with uncertainties larger than about
60\,K should be regarded as less reliable. In particular, the formal
uncertainty is listed in each case, even where it exceeds the calculated
temperature, in order to indicate the relative reliability of each
determination. We have also determined $T_{ex}$ for
each sight line from the slope in a plot of $N(J=2,3,4,5)$ versus the
excitation potential.\footnote{More generally, T$_{ex}$ is associated with
values of T$_{J-2,J} \equiv \frac{E_J - E_{J-2}}{k}
\frac{\log(e)}{\log[N(J-2)/g_{J-2}] - \log[N(J)/g_{J}]} = \frac{170 {\rm K} \times
(2J - 1)}{\ln[\frac{2J+1}{2J-3} \frac{N(J-2)}{N(J)}]}\;({\rm for}\, J \ge 2)$,
the temperature derived from comparing the populations of successive even or
odd $J$ levels.} This excitation temperature reflects the influence of
fluorescent UV pumping and H$_2$ formation in excited states on the molecular
hydrogen rotational level population \citep{jur75,bla76}. Like $T_{01}$, it is
fairly uniform among the LMC and SMC sight lines and we calculate a weighted
mean of 244$\pm$33 K, similar to previously-determined values for Magellanic
Cloud sight lines \citep{shu00}. Comparing the column density ratios
$N$(4)/$N$(2) and $N$(5)/$N$(3) for our sample with Figure 11 from
\citet{tum02}, we find that these sight lines are reasonably typical of their
high $N$(H$_2$) paths; consequently, our measurements support their conclusions
that the H$_2$ formation rate in the Clouds is low (10--40\% of Galactic rates)
and that at least some regions are illuminated by relatively intense radiation
fields (10--100 times the Galactic mean). We note, however, that the proportion
of outliers in our sample that do not fit the range of Galactic models
calculated by \citet{tum02} is much smaller than theirs. Values of $T_{01}$ and
$T_{ex}$ were also derived for the Galactic components of our H$_2$ models. A
mean of $T_{01}$ = 67$\pm$4 K was derived for the sight lines with larger
column densities and more robust measurements, although higher, and likely to
be in part non-thermal, temperatures are implied for several of the more
diffuse paths. Excitation temperatures for the Galactic components (239$\pm$38
K) are similar to those we have determined for the Magellanic Clouds.


\subsection{FUV Extinction in the Magellanic Clouds}
\label{ssection_extinction}

Before {\it FUSE}, only two reddened sight lines outside the Galaxy had been
studied in the FUV \citep{cla96}. Both regional extinction groups in the LMC
are represented by these two stars; specifically, Sk -66\degr19 is associated
with LMCAvg and Sk -69\degr270 with LMC2 (Misselt et al. 1999). The UV
extinction curves for these stars are typically non-CCM, but they share the
trait of most Galactic curves in that their UV extinction slope blends smoothly
into the FUV. These two sight lines have been included for comparison in
Table~\ref{fmtab}, which defines the LMC2 and LMCAvg subsamples, although they
were left out of the average extinction curves for those regions. \citet{hut01}
published FUV extinction curves for 3 sight lines in each of the LMC and the
SMC using {\it FUSE} data. Unfortunately, the values of $\Delta$(B-V) for the
reddened/unreddened star pairs chosen for that study are too small
($\sim$0.02-0.07 mag) for detailed extinction studies. More specifically, these
reddening values are comparable in size to the variation in the foreground
$E(\bv)$ and similar to the magnitude of the photometric uncertainties
\citep{sch91,oes95,mas02}. So, within the uncertainties there is no significant
reddening difference for any of the star pairs in \citet{hut01}, and those
sight lines are not included in this study.

The new LMC and SMC FUV extinction curves presented here seem to follow the
trend of Sk -69\degr270 and Sk -66\degr19. Despite any small remaining offsets
between the {\it IUE} and {\it FUSE} extinction curves (see
\S~\ref{ssection_h2modelling}), the new FUV curves closely follow an
extrapolation of
their FM fits from the UV. Out of our sample, only two curves exhibit
significant deviations from this overall trend, AzV 456 and Sk -69\degr228.
AzV 456 is our lone reddened star situated in the SMC wing, and the {\it FUSE}
portion of its extinction curve more closely resembles its FUV CCM curve than
an extrapolation of its FM fit in the UV. Coincidentally, the disagreement with
the FM fit begins near
the onset of molecular hydrogen absorption features; however, attempting to
reconcile the two portions of the {\it FUSE} curve by adjusting the H$_2$ column
densities indicates that a single component for the Clouds cannot reproduce the
observed continuum level shift, nor does it appear that a simple combination of
overlapping components would accomplish this result. The effect does not
appear to be related to a difference in the sensitivity of the various {\it
FUSE} channels either, since the continuum across this spectral region is
contained in both the LiF1B and LiF2A detector bands. Similar kinks are
apparent to a much smaller degree in the Galactic curves for HD 62542, HD
73882, and HD 210121 (Paper I); nevertheless, those curves match the FM fits
within the formal uncertainties [approximately 14\% in A($\lambda$)/A(V) across
the {\it FUSE} band] and we feel that it would not be appropriate to speculate
further about the origin of these features until a larger data sample is
compiled.

The FUV extinction curve for Sk -69\degr228 is very noisy, which would usually
indicate a poor match with its unreddened comparison star, but this same star
is a good match through the UV. The problem seems to arise with a complex of
photospheric \ion{Fe}{3} lines evident in the comparison star spectrum, Sk
-65\degr15. An examination of intermediate continuum segments demonstrates that
the reddened and comparison stellar spectra have very similar shapes aside from
these lines, and that connecting the lower wavenumber portion of this star's
FUV extinction curve with the two highest wavenumber points would be a
reasonable approximation. Notably, this procedure results in an FUV curve that
is consistent with an extension of the UV FM fit, despite the impression that a
small offset is still present after the {\it FUSE} flux adjustment. The curve
is still included in further analysis for the sake of completeness and because
our comparison sample is limited, but we note that the mean curve derived for
the LMC2 group, which includes this path, is consistent with the other curves
in the group unaffected by any taint of mismatch. On the
whole, the ease with which the UV and FUV extinction curves line up for our
Magellanic Cloud sight lines complements the results for Milky Way sight lines;
namely, that most Galactic sight lines also seem to show FUV extinction that is
a good extrapolation of the FM fits made to the extinction curves in the UV.
Agreement between the CCM relation and FUV extinction, however, has proved less
ubiquitous. For instance, \citet{bus94} found that the FUV extinction for two
Galactic sight lines with large ($\rho$ Oph) and small (HD 25443) values of
$R_V$ follow CCM closely, but that there are exceptions; one notable example is
the bright-nebula sight line toward HD 37903 which shows a steeper extinction
in the FUV, based on {\it Copernicus} data, than would be predicted by CCM.
Paper I, which included paths such as HD 210121 and HD 62542 that do not follow
CCM in the UV, came to similar conclusions. Although the UV extinction curves
for each sight line they studied could be smoothly extrapolated into the FUV
using an FM fit, only three paths that followed CCM through UV wavelengths were
also in accord across the FUV. The CCM relations for four others from Paper I
either over or underestimated their FUV extinction, as small discrepancies in
the UV became more pronounced at shorter wavelengths, and the two non-CCM
curves diverged more strongly from their
$R_V$-based curves with increasing wavenumber. Because the LMC and SMC curves
do not have a CCM-like wavelength dependence and because the S/N ratios of the
individual extinction curves are relatively low, subtle variations in the FUV
slope such as those seen by \citet{bus94} cannot be ruled out. 

\subsection{FUV MEM Modelling: Grain Properties}
\label{ssection_MEM}

To increase the S/N for MEM modelling purposes, we have constructed average
extinction curves for the LMC average and LMC2 regions outlined by Misselt et
al. 1999 (see Table~\ref{fmtab}); the curves are plotted in
Figure~\ref{extcurves}. The two sight lines in the SMC, which represent the bar
and wing regions, are considered individually. Using the two average LMC curves
and the two SMC curves, we have extended the dust-grain population analysis of
\citet{cla03} into the FUV. We employ a (slightly) modified version of the MEM
extinction-fitting algorithm \citep{kim94,kim96}. Instead of using the number
of grains of a given size to describe the dust population, the algorithm
employs the mass distribution in which
$m(a) da$ is the mass of dust grains per H atom in the size interval from $a$
to $a + da$. Thus, the traditional MRN-type model \citep{mat77} becomes $m(a)
\propto a^{-0.5}$. We use a power law with exponential decay (PED) as the
template function for each component. The data are examined at 34 wavelengths,
and the grain cross sections are computed over the range 0.0025--2.7 {\micron}
with 50 logarithmically spaced bins. The shape of the mass distribution is
strongly constrained only for data over the interval 0.02--1 \micron. Below
0.02 \micron, the Rayleigh scattering behavior constrains only total mass;
above 1 \micron, the "grey" nature of the dust opacity also forces the MEM
algorithm to simply adjust the total mass, using the shape of the template
function to specify the size dependence of the distribution. The total mass
of dust is constrained using both the gas-to-dust ratio and "cosmic" abundances
(i.e., we try not to use more carbon or silicon than is available). The
elemental abundance standards used in this analysis are those adopted in Paper
I for the Galaxy (358 and 35 atoms per million H for C and Si, respectively),
and by \citet{cla03} for the LMC and SMC (110 and 65 for the LMC and 54 and 11
atoms per million H for the SMC); Table~\ref{hydtab} reports the values of the
gas-to-dust ratios used here for the LMC and SMC. We consider only
three-component models of homogeneous, spherical grains: modified "astronomical
silicate" \citep{wei01}, amorphous carbon \citep{zub96}, and graphite
\citep{lao93}. While it must be acknowledged that the three grain component
system we have used to model these extinction curves is simpler than might be
expected of actual interstellar dust, the results can be quite useful for
identifying grain population properties that distinguish sight lines from one
another. The same component-specific PED constraints (e.g., the onset of the
exponential cutoff) adopted in Paper I were utilized for the current modelling.

The MEM fits to FM parameterizations of the average Magellanic Cloud extinction
curves listed in Table~\ref{fmtab} are presented in Figures~\ref{MEMfits} and
\ref{MEMmass}. The first plot shows the amount of extinction provided by the
three distinct grain components, as well as the total extinction of the model,
compared to the FM fit associated with each extinction curve. The error bars
plotted on these fits are indicative of the mean gap between the FM fit and the
underlying extinction curve in each wavelength bin. The fractions of the
adopted elemental abundances available for each grain component that are
required by the best MEM fit are listed in Table~\ref{MEMtab}.
Figure~\ref{MEMmass} depicts the corresponding mass distributions for different
sizes of grains belonging to each model component relative to the mass of
hydrogen.

The proportions of the available silicon and carbon used in the MEM fits to the
various FM parameterizations cover a very wide range, as was seen in fits to
IR through UV data alone by \citet{cla03}. Three general factors determine the
fraction of silicon and carbon that any individual sight line will use. First,
the higher the gas-to-dust ratio is, the more metals are available in the gas
phase. Second, the higher the abundances of metals are, the more material is
available. Finally, high values of $R_V$ imply a greater than average mass
fraction in larger grains which are not as efficient per unit mass as smaller
grains. For instance, it can be seen from the MEM fits that the SMC wing (AzV
456), which
has a low gas-to-dust ratio and low elemental abundances relative to the
Galaxy, uses more than 100\% of the available silicon. However, the LMC2 and
the SMC bar (AzV 18) regions, which also have both low elemental abundances but
are
characterized by higher gas-to-dust ratios, use less than half the amount of
available silicon that the LMCAvg and SMC wing regions require and also
significantly less carbon.

A recent observation has suggested that silicon is relatively undepleted in the
SMC bar \citep{wel01}. However, model fits to the average SMC bar extinction
indicate that its curve cannot be fitted with carbon grains alone
\citep{wei01,cla03}. Figures~\ref{MEMfits} and \ref{MEMmass} clearly show that
this conclusion is born out in the current fits. SMC extinction for both bar
and wing is very steep in the FUV, necessitating the presence of a large
population of small grains. Yet because the 2175 {\AA} bump is absent in the
bar region, small ($a < 0.02\micron$) carbon grains do not meet the SMC FUV
extinction requirements; thus, silicates play a very important role in the
models. Even our SMC wing sight line, whose extinction curve follows CCM
reasonably
well, places a much higher demand on silicon reserves than carbon: a typical
Galactic curve conforming to CCM allows graphite to fulfill a significant part
in FUV extinction due to the strength of its UV bump, but the SMC wing bump is
somewhat weaker than the $R_V$=2.19 CCM curve would imply. The next steepest
sight lines, in the LMC, do not show much difference in demand from a
typical Galactic curve (HD 14250 was chosen from Paper I for illustration
purposes), once the gas-to-dust ratio differences are taken into account.
Typically, carbon grains are responsible for most of the visible
extinction and silicon grains for most of the UV extinction; therefore, in
general, both species of grains are needed along any sight line to get a good
fit to the extinction curve. Apart from deficits of large silicate and carbon
grains in the SMC, the MEM analysis of the Magellanic Cloud FUV extinction
produces results similar to those derived from an analysis of several Galactic
sight lines (Paper I). The Galactic sight lines, like the SMC wing or LMCAvg,
are more likely to use more than 100\% of the available silicon and/or carbon.
Among these sight lines, the strength of the 2175 {\AA} bump in relation to the
FUV rise is the factor that distinguishes whether the stiffer abundance demand
is placed on silicon or carbon. Likewise, the element most demanded by the SMC
bar sight line and the mean LMC2 curve is determined by this factor, although
the minimal presence of the bump in these curves allows the modelling to make
more efficient use (especially in small grains) of the available elements.

The general structure in the size vs $m(a)$ curves for extinction out to 8
$\micron^{-1}$ for the Magellanic Cloud sight lines does not change much when
the {\it FUSE} data are included. Figures~\ref{MEMfits} and \ref{MEMmass}
showing the MEM fits and size vs $m(a)$ curves cannot be directly compared to
Figure 3 of \citet{cla03} because there are differences in the input parameters
such as the gas-to-dust ratios. But the MEM models run with the same input
parameters for extinction curve that are cutoff at 8 $\micron^{-1}$ and those
shown here, which extend to 10 $\micron^{-1}$, are quite similar; there is at
most a few percent difference in the amount of silicon and carbon required by
the two sets of models for the four SMC and LMC curves. The structure evident
in the mass distributions of Figure~\ref{MEMmass} particularly identify grain
sizes for which each component in our
simple model can account for the detailed shape
of each extinction curve. For instance, the two peaks in the amorphous carbon
distribution for HD 14250 correspond roughly to maxima appearing in the
silicate distribution of \citet{kim94} in their silicate-graphite model. These
authors ascribe the peaks to the requirements of fitting the optical and UV
portions of the extinction curve using the spectral wavelength dependence
appropriate to each of the two grain components in their model. In the case of
AzV 456 (the
SMC wing), however, efforts to reproduce the extinction curve using this
three-component model appear to require dramatic peaks and dips in the mass
distribution as a function of grain size. Aside from this curve, however, the
scale of structure in the distribution is consistent with levels noted by
previous studies (Kim et al. 1994; \citealt{cla03}). For AzV 456, the solution
requires that the graphite distribution be strongly peaked in order to account
for the 2175 {\AA} bump and provide as much extinction as possible at FUV
wavelengths.
Since the FUV portion of this curve extends, or even amplifies, the steep rise
at the short-wavelength end of the previous IR through UV result \citep{cla03},
there is no relaxation of the demand on silicon reserves. Specifically, when
comparing MEM solutions for the SMC wing curve with and without the {\it FUSE}
data, nearly identical proportions of the available silicon and carbon are
utilized. The only notable distinction between the two solutions is that a few
percent of the amorphous carbon demand is shifted to graphite when the FUV data
are included. The extinction curves for our other groups also smoothly extend
from the UV into FUV wavelengths, and the models derived from IR-to-UV and
IR-to-FUV data require almost precisely the same amounts of silicon and carbon
be present in dust grains. The similarity in the MEM results for these two sets
of models may
be due to the efficiency of the small grains in FUV extinction, in that large
numbers of such grains are not needed, or perhaps the observed amounts of FUV
extinction can be provided by the larger grains which are also important for
extinction at longer wavelengths. This issue will be investigated further as
we study the distinction between global and sight-line-specific FUV extinction
characteristics of using a much larger number of sight lines, in a future paper.

\section{Conclusion}
\label{section_conclusion}
This paper is the second in a series investigating the FUV characteristics of
extinction due to interstellar dust. The first paper dealt with a small sample
of Galactic sight lines with a variety of $R_V$ values, including paths whose
extinction does not conform to the CCM parameterization. This paper examined a
similarly-sized set of sight lines probing the Small and Large Magellanic
Clouds, with the goal of determining how the addition of FUV data to their
corresponding extinction curves affected models of Magellanic Cloud dust grain
properties.

Using {\it FUSE} observations of four stars in the SMC and 12 stars in the LMC,
we were able to construct two extinction curves for the SMC and seven
representing the LMC. Despite the poor quality of the data for some of the
reddened stars, molecular hydrogen was measured for each sight line and their
corresponding absorption features were removed from consideration in the
extinction analysis. Flux mismatches between {\it IUE} and {\it FUSE} data,
likely due to poor S/N values in the affected observations, were alleviated by
rescaling the entire {\it FUSE} spectrum. Comparison of the resulting FUV
extinction with the curve constructed only through wavelengths observed by {\it
IUE} demonstrated that the FUV portions were generally consistent with a smooth
extrapolation of the IR-to-UV curve. Nevertheless, we note that a ``kink''
appeared in the AzV 456 curve, similar to much smaller features present in
curves analysed in Paper I. We plan to address the existence and nature of this
kink in a subsequent paper examining the variation in properties of FUV
extinction.

It was anticipated that including FUV data in the MEM analysis of the
Magellanic Cloud dust grain population might give rise to significantly
different properties than Galactic dust since the FUV rise in the Clouds is
generally much steeper than it is in the Milky Way, and the UV bump is smaller.
However, aside from some complex structure likely driven by the simplicity of
our grain component model, the sight lines through the Clouds exhibitting
Galactic-like extinction are similar to Galactic paths in that they demand
100\% or more of the available carbon and/or silicon. The only real differences
between the populations for these galaxies are that the SMC wing sight line has
a much smaller grain size cutoff for astronomical silicates and amorphous
carbon and a generally smaller dust mass. MEM solutions for the sight lines
through the Clouds that exhibited more distinctly Magellanic-Cloud-type
extinction were able to satisfy their curves using smaller proportions of the
elements available to them; the SMC bar solution, like that for the SMC wing,
is characterized by a smaller grain size cutoff and lower dust mass than
Galactic or LMC solutions. The addition of the {\it FUSE} data to the analysis
does not dramatically alter the properties of each dust population, except to
shift small amounts of the carbon demand between amorphous carbon and graphite.

The molecular hydrogen abundances determined in the process of constructing the
full IR-to-FUV extinction curves supplement those already appearing in the
literature. Although the emphasis has been on eliminating H$_2$ absorption
features from the spectra rather than deriving robust column density
measurements, our results are consistent with the larger recent survey of
molecular hydrogen in the Magellanic Clouds by \citet{tum02} and we complement
their paths by probing several sight lines with larger column densities. Among
these sight lines, $E(\bv)$/$N$(\ion{H}{1}) ratios are generally reduced
relative to Galactic values while $E(\bv)$/$N$(H$_2$) are roughly the same,
implying somewhat lower $f$(H$_2$)-values in the Clouds than are typical in the
Milky Way for similar degrees of reddening.

\acknowledgments
This study was supported by NASA grant NAG5-108185.

\clearpage

\clearpage

\clearpage

\clearpage

\clearpage

\begin{figure}
\plotone{f5.eps}
\caption{LMC Gas-to-Dust Ratios.\newline
Gas-to-dust ratios for LMC extinction pairs are plotted above as a function of
$E(B-V)$. Of particular note, Sk -67\degr2 alone among the reddened LMC stars
is characterized by a log$_{10}$[$N$(\ion{H}{1})/$E^\prime(\bv)$] ratio
consistent with that of a Galactic star (dotted line; \citealt{dip94}), whereas
the corresponding value of log$_{10}[N{\rm (H_{Total})/}E^\prime(\bv)]$ agrees
with other LMC ratios. The LMC value for
log$_{10}$[$N$(\ion{H}{1})/$E^\prime(\bv)$] (dashed line; \citealt{koo82}) is
also plotted in both graphs for reference.
\label{gtdebvplot}}
\end{figure}

\clearpage

\begin{figure}
\epsscale{0.75}
\plotone{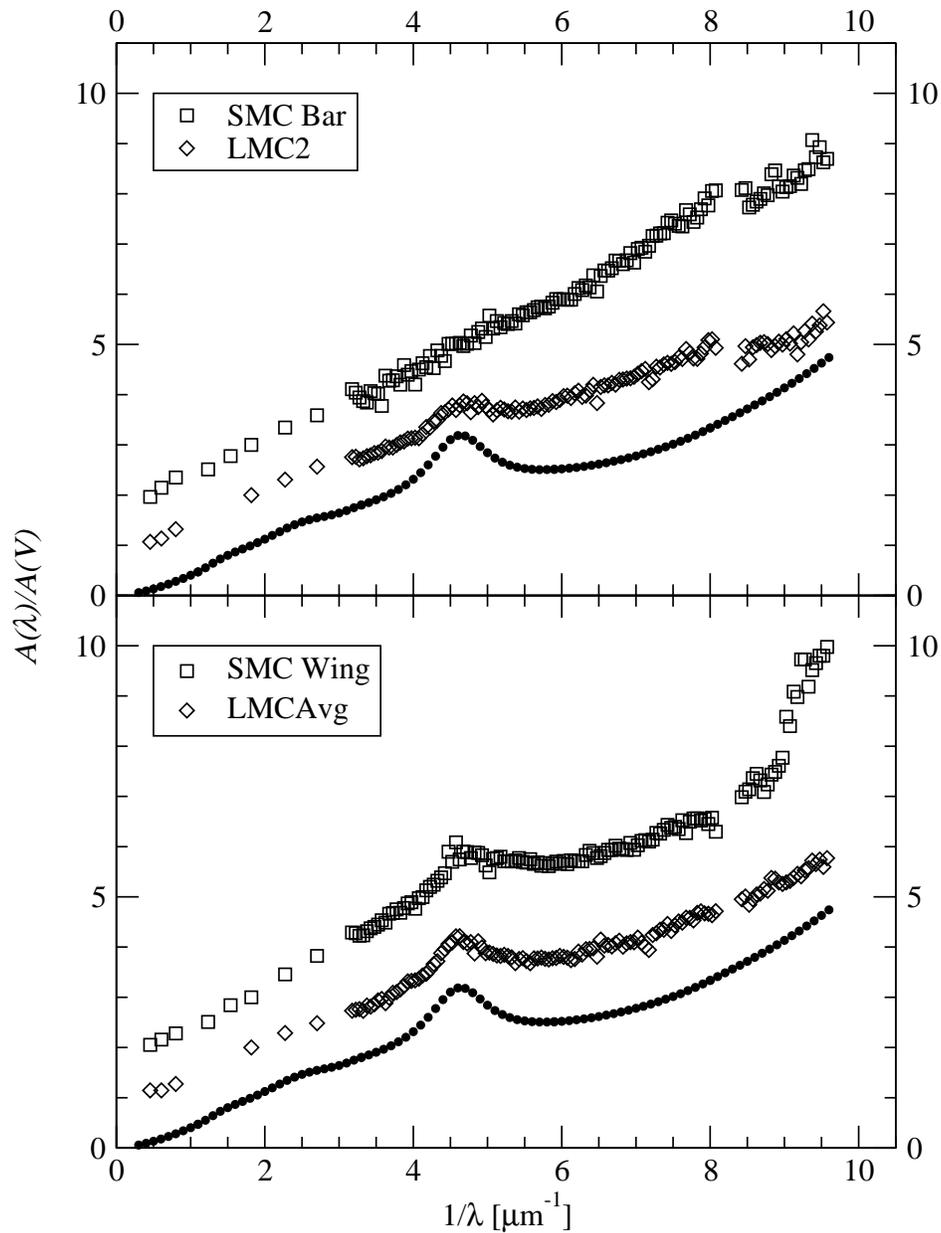}
\caption{Comparing the LMC, SMC, and Galactic Average Extinction Curves.\newline
Average extinction curves for the Magellanic Cloud regions LMC2 and SMC bar,
those more distinct in character from Galactic extinction, are shown in the
upper panel of the above plot; the lower panel depicts the mean curves for the
LMCAvg and SMC wing groupings. Both panels include the CCM $R_V$=3.1 curve in
the role of a Galactic reference. The LMC and SMC curves are offset 1 and 2
units in $A(\lambda)/A(V)$, respectively, from the Galactic curve in each panel.
\label{extcurves}}
\end{figure}
    
\clearpage

\begin{figure}
\epsscale{0.8}
\plotone{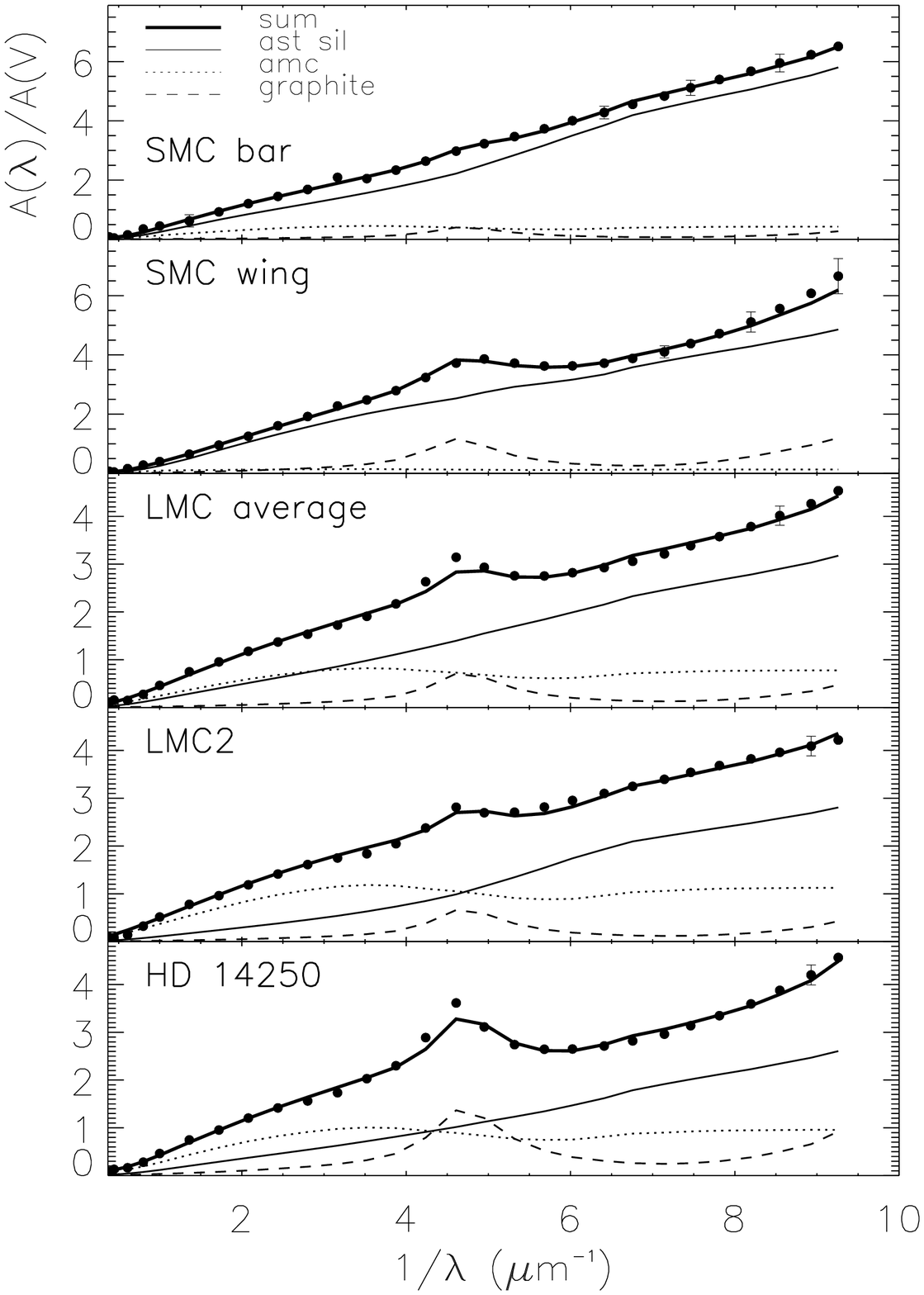}
\caption{MEM Average Extinction Curve Fits.\newline
The MEM fits of astrophysical silicate, amorphous carbon, and graphite dust
components to FM parameterizations of the average extinction curves in
Figure~\ref{extcurves} are plotted above. It should be noted that the MEM
models were adjusted to reproduce the FM fit to each extinction curve;
consequently, the SMC wing curve does not match the AzV 456/AzV 70 extinction
plot of Figure~\ref{allcurves} in full detail. The error bars are indicative of
the deviation between the FM fit and the extinction curve in nearby bins. For
comparison, a typical Galactic sight line is represented in this plot by HD
14250 ($R_V$=2.98$\pm$0.14; Paper I).
\label{MEMfits}}
\end{figure}
    
\clearpage

\begin{figure}
\epsscale{0.8}
\plotone{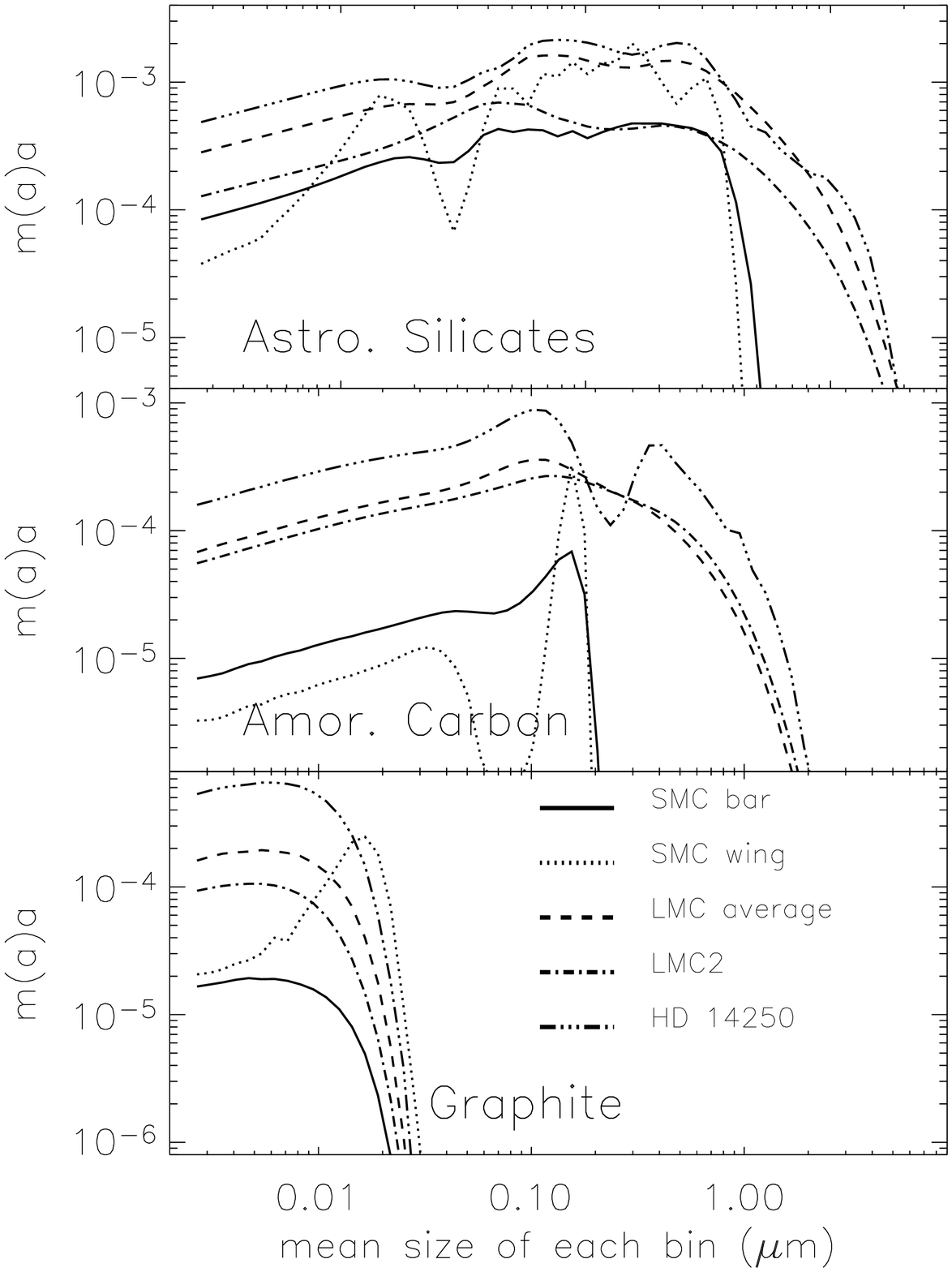}
\caption{MEM Dust Grain Mass Distributions.\newline
The MEM grain size vs $m(a) da$ functions for each of the average extinction
curves of Figure~\ref{extcurves} are plotted above. Of particular note, the
SMC grain distributions do not include silicate or amorphous carbon grains as
large as are required to fit extinction curves in the LMC or the Galaxy. Also,
the graphite grain distribution for the SMC wing is strongly peaked by the
constraints implied in the bump strength and FUV rise of its extinction curve.
As in Figure~\ref{MEMfits}, HD 14250 stands in for a typical Galactic sight
line.
\label{MEMmass}}
\end{figure}

\clearpage

\begin{deluxetable}{llcr@{$\pm$}lr@{$\pm$}lr@{$\pm$}lr@{$\pm$}l}
\tablecaption{Extinction Curve Pairs \label{exttab}}
\rotate
\tablewidth{0pt}
\tablehead{
Reddened&Comparison&Spectral&\multicolumn{2}{c}{$\Delta(\bv)$}&
\multicolumn{2}{c}{$R_V$}&\multicolumn{2}{c}{$A_V$}&
\multicolumn{2}{c}{$N($\ion{H}{1}$)/A_V$}\\
Star&Star&Type&\multicolumn{2}{c}{(mag)}&\multicolumn{2}{c}{}&
\multicolumn{2}{c}{(mag)}&
\multicolumn{2}{c}{(10$^{21}$ \ion{H}{1} atoms/$A_V$)}\\
}
\startdata
AzV 18         & AzV 462        & \ion{B2}{1}a   & 0.17 & 0.03 & 2.90 & 0.42 & 0.49 & 0.11 & \hskip 0.75cm 17.27 & 4.30 \\
AzV 456        & AzV 70         & \ion{O9.7}{1}b & 0.26 & 0.03 & 2.19 & 0.23 & 0.57 & 0.08 &  7.01 & 1.22 \\
Sk -67\degr2   & Sk -66\degr35  & \ion{B2}{1}a   & 0.15 & 0.05 & 3.75 & 0.36 & 0.56 & 0.23 &  1.78 & 0.67 \\
Sk -68\degr26  & Sk -66\degr35  & \ion{B3}{1}a   & 0.19 & 0.03 & 3.45 & 0.27 & 0.64 & 0.16 &  5.46 & 1.08 \\
Sk -68\degr129 & Sk -68\degr41  & \ion{O9}{1}a   & 0.17 & 0.05 & 3.37 & 0.29 & 0.57 & 0.22 &  6.98 & 2.36 \\
Sk -68\degr140 & Sk -68\degr41  & \ion{B0}{1}a   & 0.20 & 0.05 & 3.34 & 0.25 & 0.67 & 0.22 &  5.99 & 1.75 \\
Sk -68\degr155 & Sk -67\degr168 & \ion{O8}{1}a   & 0.20 & 0.05 & 2.81 & 0.22 & 0.56 & 0.18 &  8.88 & 9.21 \\
Sk -69\degr228 & Sk -65\degr15  & \ion{B2}{1}a   & 0.15 & 0.05 & 3.54 & 0.34 & 0.53 & 0.23 &  6.60 & 2.50 \\
Sk -69\degr279 & Sk -65\degr63  & \ion{O9}{1}a   & 0.21 & 0.05 & 3.54 & 0.25 & 0.75 & 0.23 &  5.38 & 1.50 \\
\enddata

\tablecomments{The Galactic foreground extinction is considered to be
comparable for the stars in each pair; consequently, the properties listed
in this table are appropriate only to the Magellanic Cloud dust component
\citep{mis99}. Spectral types refer to the UV classification of each star; the
sources are \citet{gor98} and \citet{mis99}.}

\end{deluxetable} 

\begin{deluxetable}{lccc}
\tablecaption{{\it FUSE} LMC and SMC Observations \label{obstab}}
\tablewidth{0pt}
\tablehead{
Sight Line&{\it FUSE} Data Set&Date&Exposure Time (s)\\
}
\startdata
AzV 18         & A1180101000 & 2000 May 29 &  9293 \\
               & B0890101000 & 2001 Jun 13 & 45245 \\
AzV 70         & A1180202000 & 2000 Oct 03 &  2831 \\
               & A1180203000 & 2000 Oct 05 &  1971 \\
               & B0900601000 & 2001 Jun 15 &  6152 \\
AzV 456        & Q1070101000 & 2000 Oct 04 &  3691 \\
               & Q1070104000 & 2000 Oct 06 &  4903 \\
               & Q1070106000 & 2000 Oct 09 &  5641 \\
               & Q1070102000 & 2000 Oct 10 &  2687 \\
               & Q1070103000 & 2000 Oct 12 &  4352 \\
               & P2210201000 & 2001 Jun 14 &  8278 \\
AzV 462        & A1180301000 & 2000 Jul 03 &  5550 \\
Sk -65\degr15  & B0861001000 & 2001 Nov 17 &  4293 \\
Sk -65\degr63  & A0490701000 & 1999 Dec 16 &  5927 \\
               & M1142001000 & 2000 Sep 26 &  5581 \\
               & B0861101000 & 2001 Sep 22 &  4125 \\
Sk -66\degr35  & B1280101000 & 2001 Oct 25 &  4030 \\
               & B0860301000 & 2002 Sep 22 &  4480 \\
Sk -67\degr2   & B0860101000 & 2001 Aug 14 &  6935 \\
Sk -67\degr168 & B0860901000 & 2001 Sep 22 &  4102 \\
Sk -68\degr26  & B0860201000 & 2001 Sep 17 & 11391 \\
Sk -68\degr129 & B0860501000 & 2001 Sep 22 &  6681 \\
Sk -68\degr140 & B0860601000 & 2001 Sep 17 & 10712 \\
Sk -68\degr155 & B0860701000 & 2001 Sep 22 &  8020 \\
Sk -69\degr228 & B0860401000 & 2001 Sep 23 &  9422 \\
Sk -69\degr279 & B0860801000 & 2001 Sep 22 &  5991 \\
\enddata

\end{deluxetable}

\begin{deluxetable}{lccccccccr@{$\pm$}lr@{$\pm$}l}
\tablecaption{Molecular Hydrogen Models: Milky Way \label{MWmolhydtab}}
\tabletypesize{\scriptsize}
\rotate
\tablewidth{0pt}
\tablehead{
&$\log_{10}N$(H$_2$)$_{MW}$&$\log_{10}N$(0)&$\log_{10}N$(1)&$\log_{10}N$(2)&
$\log_{10}N$(3)&$\log_{10}N$(4)&$\log_{10}N$(5)&$b$&
\multicolumn{2}{c}{$T_{01}$\tablenotemark{a}}&
\multicolumn{2}{c}{$T_{ex}$\tablenotemark{b}}\\
MC Star&(cm$^{-2}$)&(cm$^{-2}$)&(cm$^{-2}$)&(cm$^{-2}$)&(cm$^{-2}$)&
(cm$^{-2}$)&(cm$^{-2}$)&km s$^{-1}$&\multicolumn{2}{c}{(K)}&
\multicolumn{2}{c}{(K)}\\
}
\startdata
\multicolumn{13}{c}{SMC}\\
\tableline
AzV 18           &16.57(0.64)&16.33&16.17&14.82&14.27&     &     &
8.1$^{+2.9}_{-1.9}$& 66&192&187&296\\
AzV 70           &18.55(0.05)&18.42&17.95&15.90&15.46&14.66&     &
4.2$^{+0.9}_{-1.6}$& 52&  6&357&786\\
AzV 462          &15.92(0.16)&15.55&15.48&14.88&14.98&     &     &
5.9$^{+2.3}_{-4.9}$& 72& 34&419&223\\
\tableline
\multicolumn{13}{c}{LMC}\\
\tableline
Sk $-$65\degr15  &18.15(0.06)&17.79&17.86&16.75&16.04&     &     &
2.4$^{+1.2}_{-1.4}$& 84&  6&165&330\\
Sk $-$65\degr63  &18.00(0.07)&17.75&17.54&16.66&16.56&     &     &
1.4$^{+1.1}_{-0.4}$& 64&  8&303&263\\
Sk $-$66\degr35  &17.78(0.46)&17.24&17.60&16.34&15.74&14.08&     &
4.8$^{+2.6}_{-2.5}$&125&310&210&227\\ 
Sk $-$67\degr2   &16.83(0.36)&16.56&16.19&15.98&15.81&     &     &
6.3$^{+5.9}_{-5.2}$& 56& 90&277&197\\
Sk $-$67\degr168 &16.38(0.35)&15.60&16.25&15.24&14.74&     &     &
3.5$^{+1.2}_{-2.2}$&200& 95&200& 95\\
Sk $-$68\degr26  &16.02(0.35)&15.61&15.32&15.58&14.67&     &     &
5.4$^{+6.0}_{-4.4}$& 60& 62&354&519\\
Sk $-$68\degr41  &15.31(0.27)&14.31&15.14&14.39&14.31&     &     &
4.3$^{+4.0}_{-3.3}$&326&230&326&230\\
Sk $-$68\degr129 &16.73(0.59)&15.76&16.39&16.15&15.87&15.13&     &
2.9$^{+6.0}_{-1.9}$&228&868&409&868\\
Sk $-$68\degr140 &19.01(0.12)&18.54&18.82&17.37&16.28&14.37&14.17&
2.5$^{+2.4}_{-1.5}$&110& 42&227& 50\\
Sk $-$68\degr155 &18.35(0.50)&18.26&17.64&15.77&15.36&14.39&     &
4.6$^{+2.2}_{-3.6}$& 47&133&330&183\\
Sk $-$69\degr228 &18.00(0.20)&17.29&17.80&17.22&16.14&     &     &
2.5$^{+5.9}_{-1.5}$&167&189&224&620\\
Sk $-$69\degr279 &16.17(0.16)&15.70&15.93&14.88&14.71&     &     &
8.3$^{+5.9}_{-6.0}$&102& 59&277&308\\
\enddata

\tablenotetext{a}{The H$_2$ kinetic temperature is assumed to be equivalent to
$T_{01}$, the temperature derived from $N$($J$=0) and $N$(1) assuming a
Boltzmann distribution ($T_{01} = {{E_1 - E_0}\over{k}}
{{\log(e)}\over{\log(N(0)/g_0) - \log(N(1)/g_1)}}$).}
\tablenotetext{b}{The excitation temperature $T_{ex}$ reflects the sum of
effects such as UV photon-pumping and excited state formation, and is
identified in this table with the slope of the $\log[N(J)/g_{\rm J}]$ vs.
$E(J)$ plot for $J \geq$ 2.} \tablecomments{This table describes the Milky Way
component of the molecular hydrogen models constructed for each target star
from the {\it FUSE} data.}

\end{deluxetable} 

\begin{deluxetable}{lccccccccr@{$\pm$}lr@{$\pm$}l}
\tablecaption{Molecular Hydrogen Models: Magellanic Clouds \label{MCmolhydtab}}
\tabletypesize{\scriptsize}
\rotate
\tablewidth{0pt}
\tablehead{
&$\log_{10}N$(H$_2$)$_{MC}$&$\log_{10}N$(0)&$\log_{10}N$(1)&$\log_{10}N$(2)&
$\log_{10}N$(3)&$\log_{10}N$(4)&$\log_{10}N$(5)&$b$&
\multicolumn{2}{c}{$T_{01}$}&\multicolumn{2}{c}{$T_{ex}$}\\
MC Star&(cm$^{-2}$)&(cm$^{-2}$)&(cm$^{-2}$)&(cm$^{-2}$)&(cm$^{-2}$)&
(cm$^{-2}$)&(cm$^{-2}$)&km s$^{-1}$&\multicolumn{2}{c}{(K)}&
\multicolumn{2}{c}{(K)}\\
}
\startdata
\multicolumn{13}{c}{SMC}\\
\tableline
AzV 18           &20.36(0.07)&20.13&19.97&17.58&17.79&16.06&15.36&
8.1$^{+2.9}_{-1.9}$& 66&  9&270& 52\\
AzV 456          &20.93(0.09)&20.85&20.15&17.67&15.62&     &     &
6.0$^{+5.9}_{-5.0}$& 45&  4& 82&175\\
AzV 462          &17.65(0.13)&16.65&17.54&16.56&16.16&     &     &
2.1$^{+1.2}_{-1.1}$&\multicolumn{2}{c}{\nodata}&234&564\\
\tableline
\multicolumn{13}{c}{LMC}\\
\tableline
Sk $-$66\degr35  &19.13(0.19)&19.02&18.46&17.02&15.89&14.78&14.26&
4.4$^{+3.0}_{-3.1}$& 49& 23&302&152\\
Sk $-$67\degr2   &20.95(0.08)&20.86&20.21&18.41&16.43&15.40&     &
5.6$^{+6.0}_{-4.6}$& 46&  5&159& 75\\
Sk $-$68\degr26  &20.38(0.08)&20.20&19.90&18.50&17.76&16.43&15.88&
2.3$^{+1.0}_{-1.3}$& 59&  7&253&260\\
Sk $-$68\degr129 &20.05(0.10)&20.00&19.08&18.06&16.65&15.87&15.55&
8.0$^{+3.8}_{-4.2}$& 40& 13&309&157\\
Sk $-$68\degr140 &19.50(0.13)&19.09&19.25&18.13&17.45&15.97&15.45&
5.4$^{+2.6}_{-2.0}$& 93& 35&248& 71\\
Sk $-$68\degr155 &19.99(0.24)&19.75&19.62&17.21&16.37&15.42&14.84&
8.2$^{+3.3}_{-4.9}$& 68& 42&341&152\\
Sk $-$69\degr228 &18.70(0.14)&18.28&18.37&17.79&17.13&     &     &
3.1$^{+6.0}_{-2.1}$& 86& 37&249&555\\
Sk $-$69\degr279 &20.31(0.07)&20.10&19.90&16.46&16.11&15.34&     &
10.0$^{+6.0}_{-6.0}$& 64&  9&389&437\\
\enddata

\tablecomments{This table details the Magellanic Cloud component of the
molecular hydrogen models constructed for each target star from the {\it FUSE}
data. The temperatures $T_{01}$ and $T_{ex}$ possess the same characteristics
as those in Table~\ref{MWmolhydtab}.}

\end{deluxetable}

\begin{deluxetable}{lclc}
\tablecaption{Flux Shifts \label{fluxshifts}}
\tablewidth{0pt}
\tablehead{
Star&{\it IUE/FUSE} Flux Ratio&Star&{\it IUE/FUSE} Flux Ratio\\
}
\startdata
AzV 18          &1.10&Sk $-$67\degr168&1.05\\
AzV 70          &1.05&Sk $-$68\degr26 &0.85\\
AzV 456         &0.85&Sk $-$68\degr41 &0.95\\
AzV 462         &1.00&Sk $-$68\degr129&1.00\\
Sk $-$65\degr15 &1.20&Sk $-$68\degr140&1.10\\
Sk $-$65\degr63 &0.85&Sk $-$68\degr155&1.20\\
Sk $-$66\degr35 &1.05&Sk $-$69\degr228&1.25\\
Sk $-$67\degr2  &0.65&Sk $-$69\degr279&1.30\\
\enddata

\tablecomments{The table ratios are derived from the calibrated fluxes recorded
by each instrument in the spectral overlap from 1150--1185 {\AA}.}
    
\end{deluxetable} 

\begin{deluxetable}{lr@{$\pm$}lr@{$\pm$}lr@{$\pm$}lr@{$\pm$}lr@{$\pm$}lr@{$\pm$}l}
\tablecaption{FM Fit Parameters \label{fmtab}}
\tabletypesize{\small}
\rotate
\tablewidth{0pt}
\tablehead{
Curve&\multicolumn{2}{c}{$c_1$}&\multicolumn{2}{c}{$c_2$}&
\multicolumn{2}{c}{$c_3$}&\multicolumn{2}{c}{$c_4$}&\multicolumn{2}{c}{$x_0$}&
\multicolumn{2}{c}{$\gamma$}\\
}
\startdata
\multicolumn{13}{c}{SMC Bar}\\
\tableline
AzV 18           &$-$4.902&1.036&2.255&0.436&0.165&0.213&   0.001&0.026&
4.697&0.078&0.738&0.012\\
\tableline
\multicolumn{13}{c}{SMC Wing}\\
\tableline
AzV 456          &$-$0.419&0.130&0.908&0.094&5.026&1.625&   0.513&0.100&
4.770&0.079&1.470&0.025\\
\tableline
\multicolumn{13}{c}{LMC}\\
\tableline
Sk $-$66\degr19\tablenotemark{a}&\multicolumn{2}{c}{$-$4.66}&
\multicolumn{2}{c}{2.02}&\multicolumn{2}{c}{1.21}&\multicolumn{2}{c}{0.85}&
\multicolumn{2}{c}{4.54}&\multicolumn{2}{c}{0.73}\\
Sk $-$67\degr2   &$-$3.479&1.585&1.723&0.648&3.108&1.296&   0.881&0.313&
4.573&0.048&0.935&0.016\\
Sk $-$68\degr26  &   0.003&0.020&0.937&0.150&2.581&0.513&   0.203&0.054&
4.638&0.049&0.898&0.015\\
Sk $-$68\degr129 &$-$2.174&0.784&1.393&0.443&1.184&0.351&   0.207&0.082&
4.567&0.071&0.688&0.041\\
Average          &$-$1.704&0.253&1.268&0.041&2.902&0.187&   0.266&0.030&
4.602&0.023&0.930&0.015\\
\tableline
\multicolumn{13}{c}{LMC2}\\
\tableline
Sk $-$68\degr140 &$-$1.929&0.820&1.323&0.387&1.093&0.525&   0.161&0.055&
4.440&0.074&0.811&0.014\\
Sk $-$68\degr155 &$-$2.842&0.982&1.615&0.458&0.897&0.336&$-$0.012&0.013&
4.611&0.062&0.684&0.028\\
Sk $-$69\degr228 &$-$2.800&1.219&1.449&0.529&0.641&0.237&$-$0.418&0.164&
4.714&0.079&0.609&0.054\\
Sk $-$69\degr270\tablenotemark{a}&\multicolumn{2}{c}{$-$3.51}&
\multicolumn{2}{c}{1.52}&\multicolumn{2}{c}{0.97}&\multicolumn{2}{c}{0.24}&
\multicolumn{2}{c}{4.62}&\multicolumn{2}{c}{0.78}\\
Sk $-$69\degr279 &$-$2.696&0.857&1.364&0.364&0.853&0.232&$-$0.124&0.057&
4.603&0.077&0.664&0.047\\
Average          &$-$2.487&0.303&1.443&0.054&0.844&0.140&$-$0.055&0.016&
4.586&0.058&0.697&0.012\\
\enddata

\tablenotetext{a}{In the interest of providing additional extinction curves for
comparison, the data for sight lines previously observed by {\it Hopkins
Ultraviolet Telescope} (Sk $-$66\degr19; Sk $-$69\degr270;
\citealt{cla96,gun98}) have been included in this table.}

\end{deluxetable}

\begin{deluxetable}{lr@{$\times$}lr@{$\times$}lccr@{$\times$}lr@{$\times$}l}
\tablecaption{Properties of the Magellanic Cloud ISM\label{hydtab}}
\rotate
\tablewidth{0pt}
\tablehead{
Sight Line or&\multicolumn{2}{c}{$N$(\ion{H}{1})}&\multicolumn{2}{c}{$N$(H$_2$)}&&
$E^\prime(\bv)$\tablenotemark{a}&
\multicolumn{2}{c}{$N$(\ion{H}{1})/$E^\prime(\bv)$}&
\multicolumn{2}{c}{$N$(H$_2$)/$E^\prime(\bv)$}\\
Extinction Pair&\multicolumn{2}{c}{(cm$^{-2}$)}&
\multicolumn{2}{c}{(cm$^{-2}$)}&log$_{10} f_{\rm H_2}$&(mag)&
\multicolumn{2}{c}{(cm$^{-2}$ mag$^{-1}$)}&
\multicolumn{2}{c}{(cm$^{-2}$ mag$^{-1}$)}\\
}
\startdata
\multicolumn{11}{c}{SMC: individual targets}\\
\tableline
AzV 18           &9.0&$10^{21}$&2.3&$10^{20}$&$-$1.35&0.174&5.2&$10^{22}$&
1.3&$10^{21}$\\
AzV 456          &1.5&$10^{21}$&8.5&$10^{20}$&$-$0.28&0.332&4.5&$10^{21}$&
2.6&$10^{21}$\\
AzV 462          &6.0&$10^{20}$&4.5&$10^{17}$&$-$2.83&0.007&8.6&$10^{22}$&
6.4&$10^{19}$\\
\tableline
\multicolumn{11}{c}{SMC: extinction pairs}\\
\tableline
AzV 18/AzV 462   &8.5&$10^{21}$&2.3&$10^{20}$&$-$1.29&0.17&5.0&$10^{22}$&
1.4&$10^{21}$\\
AzV 456/AzV 070  &4.0&$10^{21}$&8.5&$10^{20}$&$-$0.53&0.26&1.5&$10^{22}$&
3.3&$10^{21}$\\
\tableline
\multicolumn{11}{c}{LMC: individual targets}\\
\tableline
Sk $-$66\degr35  &4.0&$10^{20}$&1.3&$10^{19}$&$-$1.20&0.055&7.3&$10^{21}$&
2.4&$10^{20}$\\
Sk $-$67\degr2   &1.0&$10^{21}$&8.9&$10^{20}$&$-$0.19&0.190&5.3&$10^{21}$&
4.7&$10^{21}$\\
Sk $-$68\degr26  &3.5&$10^{21}$&2.4&$10^{20}$&$-$0.92&0.181&1.9&$10^{22}$&
1.3&$10^{21}$\\
Sk $-$68\degr129 &5.2&$10^{21}$&1.1&$10^{20}$&$-$1.38&0.201&2.6&$10^{22}$&
5.5&$10^{20}$\\
Sk $-$68\degr140 &5.5&$10^{21}$&3.0&$10^{19}$&$-$1.97&0.215&2.6&$10^{22}$&
1.4&$10^{20}$\\
Sk $-$68\degr155 &4.2&$10^{21}$&9.8&$10^{19}$&$-$1.35&0.217&1.9&$10^{22}$&
4.5&$10^{20}$\\
Sk $-$69\degr228 &4.0&$10^{21}$&4.9&$10^{18}$&$-$2.61&0.145&2.8&$10^{22}$&
3.4&$10^{19}$\\
Sk $-$69\degr279\tablenotemark{b} &3.5&$10^{21}$&2.0&$10^{20}$&$-$0.98&0.221&
1.6&$10^{22}$&9.0&$10^{20}$\\
\tableline
\multicolumn{11}{c}{LMC: extinction pairs}\\
\tableline
Sk $-$66\degr19/Sk $-$69\degr83   &7.0&$10^{21}$&1.6&$10^{20}$&$-$1.64&0.25&
2.5&$10^{22}$&6.4&$10^{20}$\\
Sk $-$67\degr2/Sk $-$66\degr35    &1.0&$10^{21}$&8.9&$10^{20}$&$-$0.19&0.18&
5.6&$10^{21}$&5.0&$10^{21}$\\
Sk $-$68\degr26/Sk $-$66\degr35   &3.5&$10^{21}$&2.3&$10^{20}$&$-$0.94&0.19&
1.8&$10^{22}$&1.2&$10^{21}$\\
Sk $-$68\degr129/Sk $-$68\degr41  &4.0&$10^{21}$&1.1&$10^{20}$&$-$1.27&0.17&
2.3&$10^{22}$&6.6&$10^{20}$\\
LMCAvg curve\tablenotemark{c}&\multicolumn{2}{c}{\nodata}&
\multicolumn{2}{c}{\nodata}&$-$1.00&\nodata&1.1&$10^{22}$&5.9&$10^{20}$\\
Sk $-$68\degr140/Sk $-$68\degr41  &4.0&$10^{21}$&3.0&$10^{19}$&$-$1.83&0.20&
2.0&$10^{22}$&1.5&$10^{20}$\\
Sk $-$68\degr155/Sk $-$67\degr168 &5.0&$10^{21}$&9.8&$10^{19}$&$-$1.42&0.20&
2.5&$10^{22}$&4.9&$10^{20}$\\
Sk $-$69\degr228/Sk $-$65\degr15  &3.5&$10^{21}$&4.9&$10^{18}$&$-$2.56&0.15&
2.3&$10^{22}$&3.3&$10^{19}$\\
Sk $-$69\degr270/Sk $-$67\degr78  &3.5&$10^{21}$&0.7&$10^{20}$&$-$1.69&0.19&
1.8&$10^{22}$&3.7&$10^{20}$\\
Sk $-$69\degr279/Sk $-$65\degr63  &4.0&$10^{21}$&2.0&$10^{20}$&$-$1.04&0.21&
1.9&$10^{22}$&9.7&$10^{20}$\\
LMC2 curve\tablenotemark{c}&\multicolumn{2}{c}{\nodata}&
\multicolumn{2}{c}{\nodata}&$-$1.50&\nodata&1.9&$10^{22}$&3.1&$10^{20}$\\
\enddata

\tablenotetext{a}{As defined in the text, $E^\prime(\bv)$ refers to the portion
of $E(\bv)$ for each star arising in the Magellanic Clouds; the value listed
for each extinction pair is equivalent to the quantity $\Delta(\bv)$ in
Table~\ref{exttab}.}

\tablenotetext{b}{The atomic hydrogen column density for Sk $-$69\degr279 is
based on $N$(\ion{H}{1})$_{LMC}$ from nearby stars, since no
previously-published measurement could be found.}

\tablenotetext{c}{Gas-to-dust ratios for each of the two LMC group mean
extinction curves are included. In order to derive representative molecular
gas-to-dust ratios, an $f$(H$_2$)-value was assumed for each curve and
compared with the atomic gas-to-dust ratio determined from our averaging code;
the results of these calculations are roughly consistent with the sight line
properties in each group. $R_V$-values determined for the LMCAvg and LMC2
mean curves are $3.49\pm0.17$ and $3.24\pm0.13$, respectively.}

\tablecomments{Comparison of our H$_2$ measurements with \citet{tum02} required
estimates of $N$(\ion{H}{1})$_{LMC}$ and E$^\prime$(\bv), the Magellanic Cloud
portion of the color excess; our sources are \citet{fit85a,fit85b}. The
properties for extinction pairs studied by \citet{gun98}, Sk -66\degr19 and Sk
-69\degr270, are listed for comparison.}

\end{deluxetable}

\begin{deluxetable}{lcccc}
\tablecaption{MEM Curve Abundance Requirements \label{MEMtab}}
\tablewidth{0pt}
\tablehead{
Extinction & Silicon & \multicolumn{3}{l}{Carbon} \\
Grouping & AS & Total & AMC & Graphite \\
}
\startdata
SMC Wing & 170\% &  44\% & 15\% & 29\% \\
SMC Bar  &  78\% &  19\% & 14\% &  5\% \\
LMCAvg   &  45\% & 100\% & 77\% & 23\% \\
LMC2     &  19\% &  78\% & 66\% & 12\% \\
HD 14250 & 114\% &  75\% & 51\% & 24\% \\
\enddata

\tablecomments{Our MEM modelling reproduced curves corresponding to the FM
parameters fit to the average extinction curves for each sight line grouping
including wavenumbers from 0.455 to 9.575 $\micron^{-1}$; a typical Galactic
sight line is represented here by HD 14250 ($R_V$=2.98$\pm$0.14; Paper I). The
demands on each element are expressed as a percentage of the total interstellar
abundance for each of the astronomical silicate (AS), amorphous carbon (AMC),
and graphite grain components.}

\end{deluxetable}

\end{document}